\documentclass[12pt,a4paper,twocolumn]{aastex61}
\pdfoutput=1 %for arXiv submission
\usepackage[USenglish]{babel}
\usepackage{amsmath,amstext}
\usepackage{graphicx}
\usepackage[T1]{fontenc}
\usepackage{apjfonts} 
\usepackage[figure,figure*]{hypcap}

\begin{document}

\shorttitle{Cosmological Constraints from Galaxy Cluster Sparsity, Cluster Gas Mass Fraction and BAO}
\shortauthors{Corasaniti, Sereno \& Ettori}

\title{Cosmological Constraints from Galaxy Cluster Sparsity, Cluster Gas Mass Fraction and Baryon Acoustic Oscillations Data}

\author{Pier-Stefano Corasaniti}
\affiliation{LUTH, UMR 8102 CNRS, Observatoire de Paris, PSL Research
University, Universit{\'e} Paris Diderot, 5 Place Jules Janssen, 92190 Meudon, France}
\affiliation{Sorbonne Universit\'e, CNRS, UMR 7095, Institut d'Astrophysique de Paris, 98 bis bd Arago, 75014 Paris, France}

\author{Mauro Sereno}
\affiliation{INAF, Osservatorio di Astrofisica e Scienza dello Spazio di Bologna, via Piero Gobetti 93/3, I-40129 Bologna, Italy}
\affiliation{INFN, Sezione di Bologna, viale Berti Pichat 6/2, 40127 Bologna, Italy}

\author{Stefano Ettori}
\affiliation{INAF, Osservatorio di Astrofisica e Scienza dello Spazio di Bologna, via Piero Gobetti 93/3, I-40129 Bologna, Italy}
\affiliation{INFN, Sezione di Bologna, viale Berti Pichat 6/2, 40127 Bologna, Italy}

\correspondingauthor{Pier-Stefano Corasaniti}
\email{Pier-Stefano.Corasaniti@obspm.fr}

\begin{abstract}
In recent years, the availability of large, complete cluster samples has enabled numerous cosmological parameter inference analyses using cluster number counts. These have provided constraints on the cosmic matter density $\Omega_m$ and the amplitude of matter density fluctuations $\sigma_8$ alternative to those obtained from other standard probes. However, systematics uncertainties, such as the mass calibration bias and selection effects, may still significantly affect these data analyses. Hence, it is timely to explore other proxies of galaxy cluster cosmology that can provide cosmological constraints complementary to those obtained from cluster number counts. Here, we use measurements of the cluster sparsity from weak lensing mass estimates of the LC$^2$-{\it single} and HSC-XXL cluster catalogs to infer constraints on a flat $\Lambda$CDM model. The cluster sparsity has the advantage of being insensitive to selection and mass calibration bias. On the other hand, it primarily constrains a degenerate combination of $\Omega_m$ and $\sigma_8$ (along approximately constant curves of $S_8=\sigma_8\sqrt{\Omega_m/0.3}$), and to less extent the reduced Hubble parameter $h$. Hence, in order to break the internal parameter degeneracies we perform a combined likelihood analysis of cluster sparsities with cluster gas mass fraction measurements and BAO data. We find marginal constraints that are competitive with those from other standard cosmic probes: $\Omega_m=0.316\pm 0.013$, $\sigma_8=0.757\pm 0.067$ (corresponding to $S_8=0.776\pm 0.064$) and $h=0.696\pm 0.017$ at $1\sigma$. Moreover, assuming a conservative Gaussian prior on the mass bias of gas mass fraction data, we find a lower limit on the gas depletion factor $Y_{b,500c}\gtrsim 0.89$.
\end{abstract}


\section{Introduction}
\label{sec:intro}
There is a wide spread consensus that observations of galaxy clusters can provide a wealth of cosmological information \citep[see e.g.][for a review]{2011ARA&A..49..409A,2012ARAA..50..353K}. In recent years the potential to probe cosmology with galaxy clusters has been explored thanks to numerous survey programs. These have provided increasingly large cluster datasets from X-ray observations of the intra-cluster gas \citep[see e.g.][]{2010MNRAS.407...83E,2016AA...592A...1P,2017AJ....153..220B}, the detection of the Sunyaev-Zeldovich (SZ) effect in the Cosmic Microwave Background (CMB) radiation \citep[see e.g.][]{2011ApJ...737...61M,2014AA...571A..29P,2015ApJS..216...27B,2016AA...594A..27P} and measurements of galaxy overdensities \citep{2016ApJS..224....1R,2019MNRAS.485..498}. 

The availability of complete cluster samples has enabled cosmological parameter inference analyses from cluster number count measurements \citep{2016ApJ...832...95D,2016AA...594A..24P,2017MNRAS.471.1370S,2018AA...620A..10P,2019ApJ...878...55B,2020PhRvD.102b3509A}. Quite remarkably, these studies have consistently found values of the amplitude of linear matter density fluctuations on the $8$ Mpc $h^{-1}$ scale $\sigma_8$ ($h$ being the reduced Hubble parameter) and the cosmic matter density $\Omega_m$ which differs from that of the {\it Planck} primary CMB analysis \citep{2016AA...594A..13P,2020A&A...641A...6P}. 
As an example, the analyses of the SZ cluster counts from {\it Planck} \citep{2014A&A...571A..20P,2016AA...594A..24P} and the South Pole Telescope \citep[SPT,][]{2019ApJ...878...55B}, as well as the cluster counts from the Dark Energy Survey Year 1 \citep[DES-Y1,][]{2020PhRvD.102b3509A} have resulted in lower values of $S_8=\sigma_8\sqrt{\Omega_m/0.3}$ compared to those from the {\it Planck} primary CMB. Similar results have been obtained using measurements of galaxy clustering from gravitational lensing shear data \citep[see e.g.][]{2017MNRAS.465.1454H,2018MNRAS.474.4894J,2018PhRvD..98d3528T,2020A&A...638L...1J}. However, in the case of cluster number counts the tension may result from systematic effects, since the statistical significance of the discrepancy depends on the amplitude of the cluster mass calibration bias \citep[see e.g.][]{2016AA...594A..24P,2019AA...626A..27S}. This is because the masses of the {\it Planck}-SZ cluster have been estimated using scaling relations calibrated on X-ray mass estimates. The latter are derived under the hydrostatic equilibrium (HE) hypothesis, consequently any departure from the HE can introduce a systematic bias which propagates into the cosmological parameter inference analysis. However, the level of mass bias necessary to reconcile the {\it Planck}-SZ cluster counts with the {\it Planck} primary CMB results contrasts with expectation from hydrodynamical simulations of galaxy clusters \citep[see e.g.][]{2012NJPh...14e5018R,2013ApJ...777..151L,2016ApJ...827..112B,2020arXiv200111508B} as well as the bias estimated from the analyses of clusters for which accurate lensing or X-ray data are available \citep{2015MNRAS.449..685H,2017MNRAS.472.1946S,2019AA...621A..40E,2019MNRAS.489..401Z,2020PASJ...72...26M}. Departures from the standard cosmological scenario can also account for such discrepancies. As an example, the presence of massive neutrinos has been shown to alleviate the tension \citep{2018AA...614A..13S}. Similarly, selection effects cannot be {\it a priori} excluded \citep[see e.g.][for a study of the impact of mass bias in weak lensing shear selected cluster samples]{2020ApJ...891..139C}. Hence, it is timely to investigate other galaxy cluster observables that can provide model parameter constraints alternative to those inferred from cluster number counts.

Cosmological information is encoded in the internal mass distribution of clusters. This is because the massive dark matter halos which host these structures have assembled through a hierarchical process that depends on the cosmic matter content, expansion rate, and the amplitude of initial matter density fluctuations. The analysis of N-body simulations has shown that the density profile of halos is described by the Navarro-Frenk-White formula \citep[NFW,][]{1997ApJ...490..493N}, such that for a halo of given mass $M$ its profile only depends on the concentration parameter $c$. Numerical studies have subsequently shown that the median halo concentration as function of the halo mass depends on the specificities of the simulated cosmological model \citep[see e.g.][]{2001MNRAS.321..559B,2003ApJ...597L...9Z,2004A&A...416..853,2009ApJ...707..354Z,2012MNRAS.422..185G}. This has suggested that measurements of the halo concentration from observations of a sample of galaxy clusters can provide cosmological constraints. However, the use of the $c-M$ relation as cosmological proxy suffers of several drawbacks. Firstly, astrophysical processes affecting the baryon distribution in the inner region of clusters may alter the estimated concentration-mass relation, thus inducing a systematic error in the cosmological analysis \citep{2008MNRAS.390L..64D,2010MNRAS.406..434M,2011MNRAS.416.2539K}. Secondly, selection effects may have a strong impact on the cosmological parameter inference \citep{2015MNRAS.449.2024S}. Hence, using measurements of the concentration of galaxy clusters to test cosmology has proven to be very challenging \citep[see e.g.][]{2010A&A...524A..68E}.

 Alternatively, \citet{2014MNRAS.437.2328B} have proposed that the sparsity, i.e. the ratio of the halo mass within radii enclosing different overdensities, can provide a non-parametric characterisation of the mass distribution in halos, while retaining the cosmological information encoded in the halo mass profile. More recently, \citet{2018ApJ...862...40C} have shown that the halo sparsity provide a cosmological proxy that is insensitive to selection and mass calibration bias. Furthermore, as the halo sparsity quantifies the excess of mass within a spherical shell comprised between two radii relative to the mass enclosed in the inner one, it is sensitive probe of the screening mechanism in Modified Gravity scenarios \citep{2020PhRvD.102d3501C}.

Here, we infer cosmological parameter constraints on $\Omega_m$, $\sigma_8$, and $h$ of a flat $\Lambda$CDM model using estimates of the average sparsity of clusters from weak lensing cluster mass in combination with cosmic distance measurements from Baryon Acoustic Oscillations (BAO) and cluster gas mass fraction from archival data.

The paper is organised as the following. In Sec.~\ref{method} we describe the methodology, in Sec.~\ref{data} the datasets used and in Sec.~\ref{results} we present the results. In Sec.~\ref{conclusions} we discuss the conclusions.

\section{Methodolgy}\label{method}
\subsection{Halo Sparsity}
Originally introduced by \citet{2014MNRAS.437.2328B}, the halo sparsity provides a non-parametric characterisation of the mass profile of halos in terms of the ratio of masses within radii enclosing two different overdensities (in units of the critical density or the background density), namely:
\begin{equation}
s_{\Delta_1,\Delta_2}=\frac{M_{\Delta_1}}{M_{\Delta_2}},
\end{equation}
where $M_{\Delta_1}$ and $M_{\Delta_2}$ are the halo masses at overdensities $\Delta_1$ and $\Delta_2$ respectively, with $\Delta_1 < \Delta_2$. As shown in \citet{2014MNRAS.437.2328B}, this provides an observational proxy of the cosmological information encoded in the halo mass profile. This is because the mass distribution within halos is the result of the hierarchical halo assembly process that depends on the growth and initial amplitude of the matter density fluctuations which in turn depend on the cosmological parameters. 

The analysis of N-body simulations has shown that the halo sparsity varies weakly with halo mass. This has very important consequences, since it implies that the halo ensemble average value of the sparsity at a given redshift and for a given cosmology can be predicted from the halo mass function at the overdensities of interest as given by
\citep{2014MNRAS.437.2328B}:
\begin{widetext}
\begin{equation}\label{spars_pred}
\int_{M_{\Delta_2}^{\rm min}}^{M_{\Delta_2}^{\rm max}}\frac{dn}{d M_{\Delta_2}}d\ln{M_{\Delta_2}}=\langle s_{\Delta_1,\Delta_2}\rangle\int_{\langle s_{\Delta_1,\Delta_2}\rangle M_{\Delta_2}^{\rm min}}^{\langle s_{\Delta_1,\Delta_2}\rangle M_{\Delta_2}^{\rm max}}\frac{dn}{d M_{\Delta_1}}d\ln{M_{\Delta_1}},
\end{equation}
\end{widetext}
this can be solved numerically for $\langle s_{\Delta_1,\Delta_2}\rangle$ given the functional form of the halo mass functions $dn/dM_{\Delta_1}$ and $dn/dM_{\Delta_2}$ respectively. The validity of Eq.~(\ref{spars_pred}) has been extensively tested using halo catalogs from high-resolution large volume N-body simulations in \citet{2018ApJ...862...40C,2019MNRAS.487.4382C} and we refer the readers to these publications for further details. In particular, for the halo masses used in this work, namely $M_{200c}$ and $M_{500c}$, which correspond to masses within radii enclosing the overdensity $\Delta=200$ and $500$ respectively (in units of the critical density $\rho_c$), the analysis of the N-body halo catalogs has shown that Eq.~(\ref{spars_pred}) holds valid to better than percent level \citep[see e.g. Table 1 in][]{2019MNRAS.487.4382C}. Hence, Eq.~(\ref{spars_pred}) sets a quantitative framework to perform cosmological parameter inference analyses using measurements of the average galaxy cluster sparsity.

Here, we predict the average halo sparsity for a given cosmology at a given redshift by solving Eq.~(\ref{spars_pred}). In the following, we assume a parametrisation of the halo mass function as given by the Sheth-Tormen formula \citep{1999MNRAS.308..119S}:
\begin{equation}
\frac{dn}{dM}=\frac{\rho_m}{M}\left(-\frac{1}{\sigma}\frac{d\sigma}{dM}\right)A\frac{\delta_c}{\sigma}\sqrt{\frac{2a}{\pi}}\left[1+\left(a\frac{\delta_c^2}{\sigma^2}\right)^{-p}\right] e^{-\frac{a\delta_c^2}{2\sigma^2}},
\end{equation}
where $\rho_m$ is the mean matter density, $\delta_c$ is the linearly extrapolated spherical collapse threshold which we compute using the formula by \citet{1996MNRAS.280..638K}, and
\begin{equation}
\sigma^2(M,z)=\frac{1}{2\pi}\int dk\, k^2 \,\tilde{W}^2[kR(M)] P(k,z),
\end{equation}
is the variance of the linear density field smoothed on a spherical volume of radius $R$ enclosing the mass $M=4/3\pi\rho_m R^3$ with
\begin{equation}
\tilde{W}[k R]=\frac{3}{(kR)^3}[\sin(kR)-kR \cos(kR)],
\end{equation}
and $P(k,z)$ being the linear matter power spectrum at a given redshift $z$, which we compute assuming the linear transfer function from \citet{1998ApJ...496..605E}. Notice that due to the exponential cut-off of the halo mass function at large masses, the upper limits in the integrals of Eq.~(\ref{spars_pred}) can be set to an arbitrarily large value without affecting the evaluation. Thus, we have set $M^{\rm max}_{500c}=10^{16}\,M_{\odot}\,h^{-1}$. In the case of the lower limits, we have set $M^{\rm min}_{500c}=10^{13}\,M_{\odot}\,h^{-1}$ consistently with the lowest mass in our cluster catalogs. However, we have found that the solution of Eq.~(\ref{spars_pred}) is insensitive to the specific choice of $M^{\rm min}_{500c}$ due the weak dependence of the halo sparsity on the halo mass.

In order to account for the redshift and cosmology dependence of the mass function at the overdensities of interest, we adopt the parametrisation of the ST coefficients $a$, $p$ and $A$ given by Eq.~(12) in \citet{2016MNRAS.456.2486D}, to which we refers as ST-Despali. As shown in \citet{2018ApJ...862...40C}, the average sparsity depends primarily on $\sigma_8$ and $\Omega_m$, which are highly degenerate, and to less extent on $h$.

The computation of the average halo sparsity has been performed using a modified version of the numerical code \texttt{Halo\_Sparsity}\footnote{The code \texttt{Halo\_Sparsity} is publicly available at \url{https://github.com/pierste75/Halo_Sparsity}.} \citep{Halo_Sparsity}, which we have specifically developed for the work presented here. 

\subsection{Sparsity Systematic Errors}
\subsubsection{Halo Mass Function Parametrisation}
Using the N-body halo catalogs from the RayGalGroupSims simulation of a $\Lambda$CDM model, \citet{2018ApJ...862...40C} have shown that ST-Despali parametrisation reproduces the average halo sparsity to better than a few percent accuracy in the redshift range $0.5\lesssim z\lesssim 1.2$, while at lower and higher redshifts differences increase up to $\sim 15\%$. In the case of the \citet{2008ApJ...688..709T} parametrisation of the halo mass function deviations from the N-body simulation results are much larger and we do not consider it here. Instead, a parametrisation of the ST coefficients on the RayGalGroupSims halo catalogs \citep[see Eq.~(A4), (A5) in][]{2018ApJ...862...40C} gives average sparsity estimates that reproduce the N-body simulation results to sub-percent level at all redshifts. We will refer to this parametrisation as ST-RayGal. 

As discussed in \citet[][]{2018ApJ...862...40C}, the systematic deviation from the ST-Despali parametrisation is two-fold.  On the one hand, the ST-Despali parametrisation has been calibrated on halo catalogs from a suite of cosmological simulations which covers a volume that is eight time smaller than that of the RayGalGroupSims, thus resulting in a less accurate determination of the mass function at the high-mass end. On the other hand, the calibration has been realised using halo catalogs which contain different halos at different overdensities, rather than the same halos with masses estimated at different overdensities, as expected from Eq.~(\ref{spars_pred}). Since modelling errors on the average sparsity prediction due to the halo mass function can induce a systematic bias on the cosmological parameter analysis \citep[see Section 3.1 in][]{2018ApJ...862...40C}, we correct the ST-Despali prediction using the results from the halo catalogs of the RayGalGroupSims simulation (see Appendix~\ref{app1}). 
%Moreover, in order test the stability of the results with respect to the corrected ST-Despali prediction, we also consider the case of the ST-RayGal parametrisation. As the latter has been calibrated on a single cosmological model simulation, rather than a the suite of simulations with different cosmological model parameters, we expect to find less stringent constraints than the ST-Despali case. 

\subsubsection{Radial Dependent Halo Mass Bias}
The halo sparsity, being a mass ratio, is by definition exempt of constant mass calibration bias. Furthermore, being nearly independent of the halo mass, it is also insensitive to selection effects as shown in \citet[][]{2018ApJ...862...40C}. In contrast, the presence of a radial dependent mass bias can introduce a systematic error. As an example, astrophysical processes can alter the inner region of the dark matter halo profile, while leaving unaltered the external regions. This can induce a radial dependent bias on the mass estimation. In \citet[][]{2018ApJ...862...40C} this has been investigated in the case of hydrostatic masses. Here, we focus instead on weak lensing mass measurements which differently from X-ray observations probe the outer regions of the dark matter distribution in clusters. We estimate the level of bias that such an effect induce on the sparsity $s_{200,500}$ using the mass bias estimates at $M_{200c}$ and $M_{500c}$ obtained in \citet[][see their Fig. 2]{2018MNRAS.479..890L} for halos with $M_{200c}\ge 10^{14}$ $M_{\odot}$ from the catalogs of the Cosmo-OWLS simulations \citep{2014MNRAS.441.1270L}. These are N-body/hydro simulations that account for astrophysical feedback from supernovae and active galactic nuclei. In the case of an extreme astrophysical feedback model \citep[AGN 8.7 in][]{2014MNRAS.441.1270L}, we find that the relative variation of the halo sparsity varies from $\sim 5\%$ at the low-mass end ($M_{200c}\sim 10^{14}$ $M_{\odot}$) to $\sim 3\%$ at high-mass end ($M_{200c}\sim 10^{15}$ $M_{\odot}$). In the case of a more realistic feedback scenario for which the profile of the simulated clusters reproduces that of the observed X-ray clusters \citep[AGN 8.0 in][]{2014MNRAS.441.1270L} we find a smaller effect with a relative variation limited to $\lesssim 3\%$ over the same mass interval.

Another source of radial dependent mass bias in weak lensing observations results from fitting a spherically symmetric NFW profile \citep[][]{1997ApJ...490..493N} to tangential shear profile measurements. Deviations from sphericity of the halo mass distribution and projection effects may induce systematic errors on the estimated mass at different radii. This has been studied in \citet{2011ApJ...740...25B}. Using the mass bias estimates at $\Delta=200$ and $500$ (in units of the critical density) quoted in their Table 3 and 4, we find a relative variation of the sparsity $\Delta{s_{200,500}}/s_{200,500}\lesssim 3\%$ at $z=0.25$ and $\lesssim 2\%$ at $z=0.5$. 

All these effects are smaller compared to the systematic error due to the modelling of the mass function previously discussed, and much smaller than current uncertainties on weak lensing mass determinations from shear profile measurement errors (see Section~\ref{WLmasses}). These sources of bias are also small when compared to the intrinsic scatter in weak lensing mass estimates, which is of order of $20-40\%$ per cent per cluster \citep{2011ApJ...740...25B,2015MNRAS.450.3633S}.

\subsection{Gas Mass Fraction}
The halo sparsity is primarily sensitive to a degenerate combination of $\sigma_8$ and $\Omega_m$, while it has a weaker dependence on $h$ \citep[][]{2018ApJ...862...40C}. Hence, estimates of the cluster gas mass fraction, $f_{\rm gas}$, from X-ray observations of galaxy clusters can break the internal parameter degeneracies. This is because $f_{\rm gas}$ probes the cosmic baryon fraction, i.e. the ratio of the mean baryon density $\Omega_b$  to the mean matter density $\Omega_m$, $f_b=\Omega_b/\Omega_m$, as well as the angular diameter distance $D_A(z)$ \citep[see e.g.][]{1993Natur.366..429W,2003AA...398..879E,2004MNRAS.353..457A,2009AA...501...61E}.

Following \citet{2008MNRAS.383..879A}, we model the relation between the gas mass fraction $f_{{\rm gas},\Delta}$ estimated within a radius enclosing a given overdensity $\Delta$ and the baryon fraction $f_{\rm bar}$ at a given redshift as
\begin{equation}
f_{{\rm gas},\Delta} = K_{\Delta}\, Y_{b,\Delta} \frac{\Omega_b}{\Omega_m}\left[\frac{D_A^{\rm fid}(z)}{D_A(z)}\right]^{3/2}-f_{*,\Delta},\label{fgas}
\end{equation}
where $D_A$ is the angular diameter distance, $D^{\rm fid}_A$ is the angular diameter distance of the fiducial cosmological model assumed to infer the gas mass fraction measurement. This derives from the fact that we focus on X-ray based estimates such that $f_{{\rm gas},\Delta}=M^{\rm gas}_{\Delta}/M^{\rm tot}_{\Delta}$, where $M^{\rm tot}_{\Delta}=M^{\rm HE}_{\Delta}/K_{\Delta}$ amd $M^{\rm HE}_{\Delta}$ is the cluster mass obtained assuming hydrostatic equilibrium. Hence, differently from the halo sparsity, $f_{{\rm gas},\Delta}$ also depends on uncertain quantities that parametrise the baryonic content of clusters, such as the mass calibration bias $K_{\Delta}\equiv 1-b_{\Delta}=M^{\rm HE}_{\Delta}/M^{\rm tot}_{\Delta}$, the gas depletion factor $Y_{b,\Delta}$, and the stellar fraction $f_{*,\Delta}$\footnote{Here, we assume that $f_{*,\Delta}$ is independent of the cluster mass.}. These play the role of nuisance parameters. To be as conservative as possible, we have tested the stability of our results for different assumptions on these parameters.
 
\subsection{BAO Cosmic Distance}
Cosmic distance measurements from BAO analyses provide cosmological parameter constraints which are complementary to those inferred from estimates of the cluster sparsity and gas mass fraction. Here, we focus on measurements of the spherically-averaged cosmic distance relative to the sound-horizon at the drag epoch at different redshifts, $D_V(z)/r_d$. The spherically-averaged cosmic distance is defined as
\begin{equation}
D_V(z)=\left[\frac{c z}{H(z)}D^2_M(z)\right]^{1/3},
\end{equation}
where $c$ is the speed of light, $H(z)$ is the Hubble function, and in a flat-universe
\begin{equation}
D_M(z)=\int_0^z \frac{c dz'}{H(z')}.
\end{equation}
We approximate the cosmological dependence of the sound-horizon at the drag epoch as \citep[see e.g.][]{2015PhRvD..92l3516A}
\begin{equation}
r_d\approx \frac{55.154\, e^{-72.3\cdot0.0006^2}}{\omega_m^{0.25351}\omega_b^{0.12807}},
\end{equation}
with $\omega_{m,b}=\Omega_{m,b}h^2$. 
 
 \section{Datasets}\label{data}
 \subsection{Weak Lensing Cluster Mass Measurements}\label{WLmasses}
We estimate the sparsity of galaxy clusters using lensing mass measurements from publicly available cluster catalogs at $M_{500c}$ and $M_{200c}$. In particular, we consider a selected sample of clusters from the Literature Catalogs of Lensing Clusters \citep[LC$^2$,][]{2015MNRAS.450.3665S}, that includes as subsets the PSZ2Lens \citep[][]{2017MNRAS.472.1946S} and the Cluster Lensing and Supernova Hubble (CLASH) project \citep[][]{2012ApJS..199...25P} catalogs. In addition, we consider the Subaru Hyper Suprime-Cam (HSC) lensing mass determinations \citep{2020ApJ...890..148U} of an X-ray selected cluster sample from the XXL-XMM survey \citep{2016AA...592A...1P}, which we refer to as HSC-XXL sample.

The LC$^2$-{\it single} cluster catalog consists of weak lensing cluster masses from archival data; at the time of this analysis the latest version\footnote{We use the version 3.8 publicly available at \url{http://pico.oabo.inaf.it/~sereno/CoMaLit/LC2/}} included $672$ entries at redshifts $z<1.7$. We specifically focus on a selected subset of $187$ clusters for which cluster masses at $M_{200c}$ and $M_{500c}$ have been inferred from a two-parameter fit of the shear lensing profile\footnote{For the clusters analysed in \citet{2019MNRAS.488.1704K}, we considered the masses from the 2-parameter fits reported among the others in LC$^2$-\text{\it all}, rather than the 1-parameter fit results listed in LC$^2$-\text{\it single}.}. This is because mass estimates from a one parameter fit, such a those obtained assuming the Singular Isothermal Sphere (SIS) or the NFW profile with a fixed concentration-mass relation result in a biased determination of the halo sparsity (see Appendix~\ref{app2}). Out of this dataset, we discard $6$ clusters which we found to be outliers in the distribution of the sparsity of clusters within the same redshift bin (see Appendix~\ref{app3}). Hereafter, we will refer to this subsample as {\it Selected} LC$^2$-{\it single} clusters containing $181$ entries. This includes mass estimates from the PSZ2Lens catalog \citep{2017MNRAS.472.1946S}, that is a statistically complete and homogeneous sample of $35$ galaxy clusters at $z<0.7$ from the second {\it Planck} Catalogue of Sunyaev-Zeldovich Sources \citep[PSZ2Union,][]{2016AA...594A..27P} with weak lensing data from the CFHTLenS \citep[Canada France Hawaii Telescope Lensing Survey,][]{2012MNRAS.427..146H} and the RCSLenS \citep[Red Cluster Sequence Lensing Survey,][]{2016MNRAS.463..635H} surveys. The HSC-XXL lensing mass catalog \citep{2020ApJ...890..148U} consists of $136$ X-ray selected clusters in the redshift range $0.031 < z < 1.033$. Therefore our total sample contains $317$ clusters.

\begin{figure}[t]
\centering
\plotone{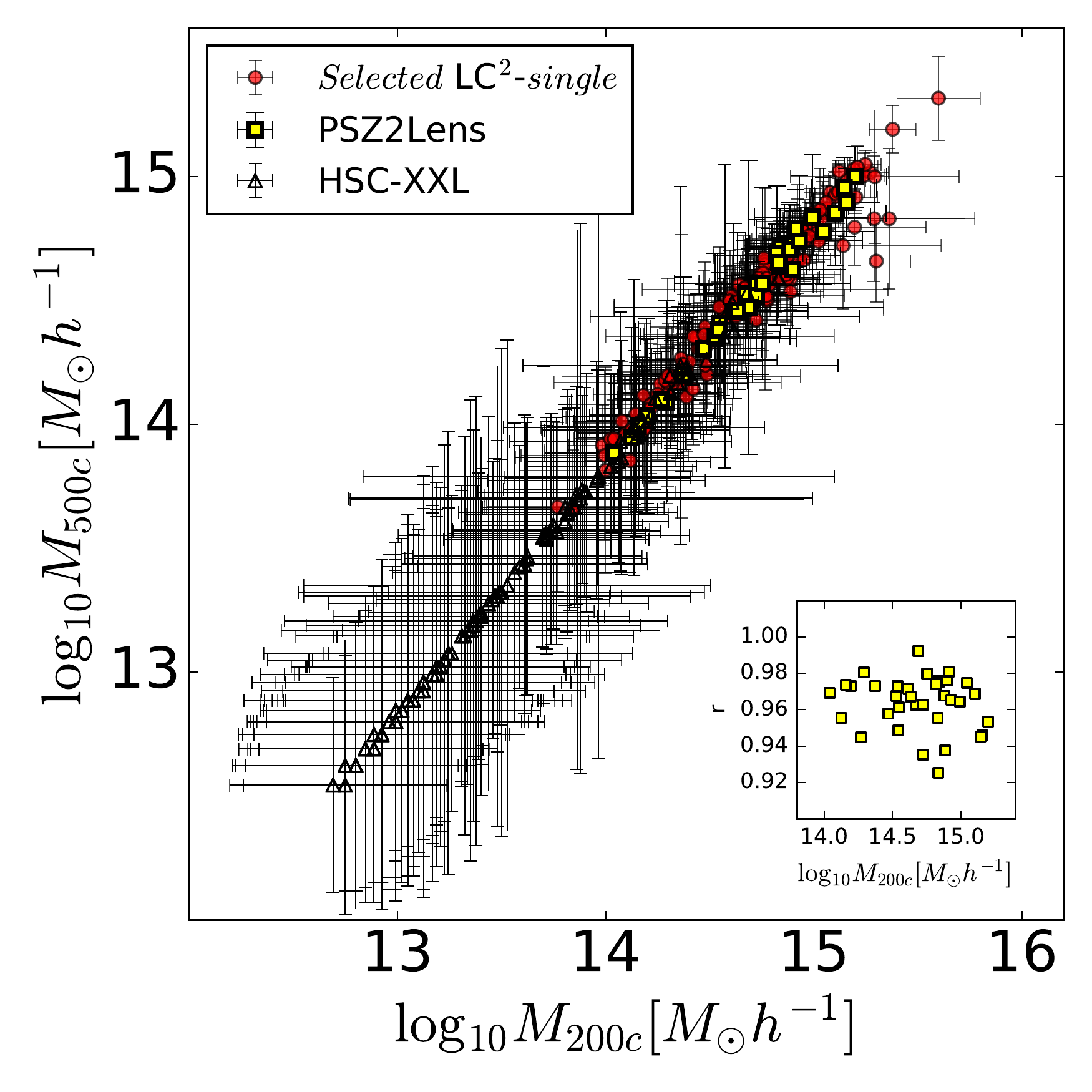}
\caption{$M_{500c}$ vs $M_{200c}$ for the {\it Selected} LC$^2$-{\it single} sample (red circles) and the HSC-XXL clusters (empty triangles). The values of the correlation coefficients of the PSZ2Lens cluster masses (yellow squares) are shown in the inset plot. \label{fig:psz2lens}}
\end{figure}

\begin{figure}[t]
\centering
\epsscale{1.1}
\plotone{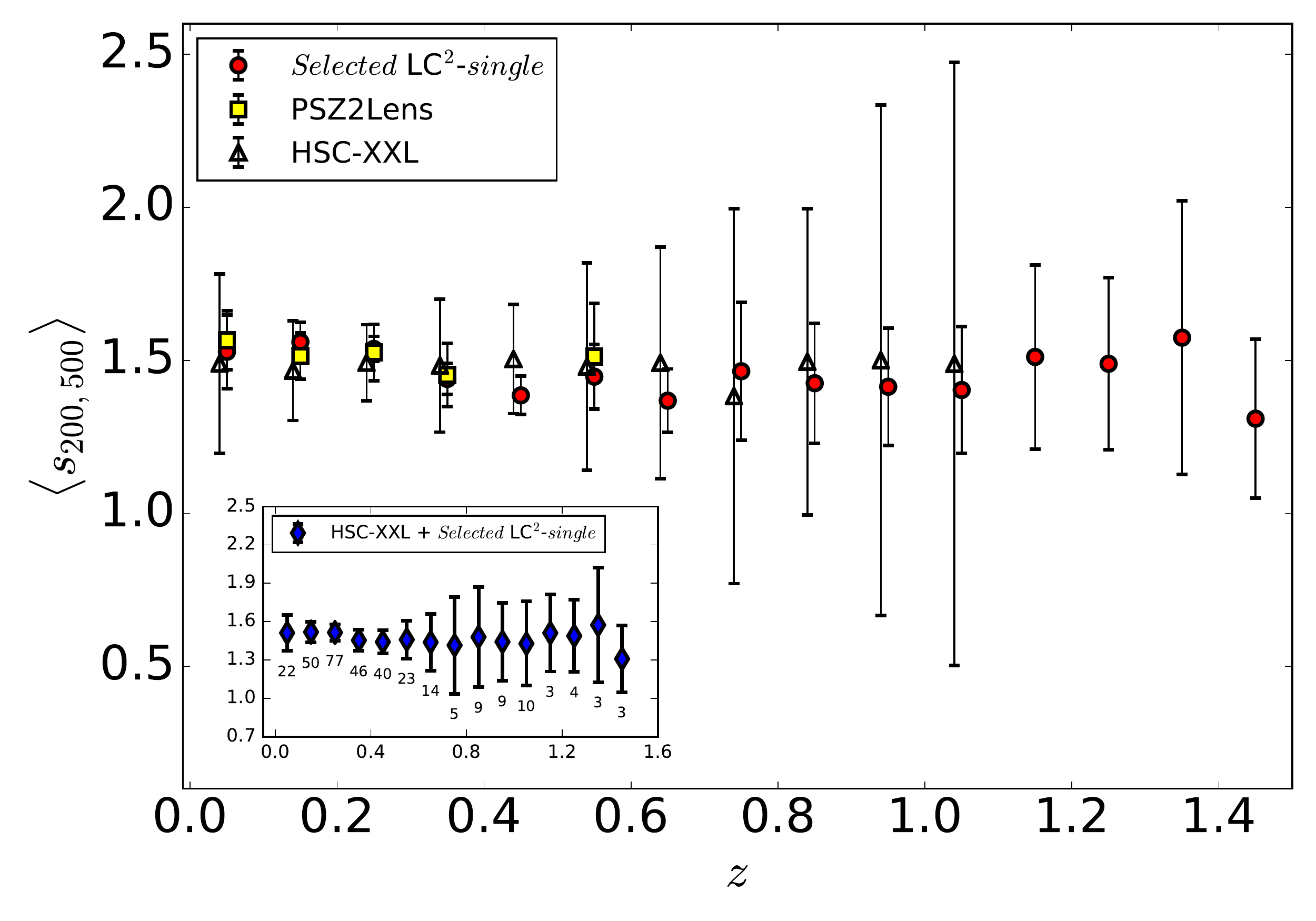} 
\caption{Average cluster sparsity in redshift bins of size $\Delta{z}=0.1$ for {\it Selected} LC$^2$-{\it single} (red points), PSZ2Lens (yellow squares), and HSC-XXL (empty triangles) clusters. The average sparsity estimates for the combined sample {\it Selected} LC$^2$-{\it single} and HSC-XXL (blue circles) are shown in the inset plot, the number of clusters in each bin is given below every data point.\label{fig:s200500binned}} 
\end{figure}

In Fig.~\ref{fig:psz2lens} we plot $M_{200c}$ vs $M_{500c}$ for the clusters in the {\it Selected} LC$^2$-{\it single} sample, the PSZ2Lens sub-sample, and the HSC-XXL dataset respectively. Estimates of the correlation coefficient between the two mass measurements are only available for the PSZ2Lens sub-sample, these are shown in the inset plot. We see that, being the sparsity very weakly dependent on selection effects, data from homogeneous samples such as PSZ2Lens or HSC-XXL are in very good agreement with the heterogeneous {\it Selected} LC$^2$-{\it single} clusters. This appears more clearly in Fig.~\ref{fig:s200500binned}, where we have plotted the average sparsity for the different samples in redshift bins of size $\Delta{z}=0.1$ containing at least $2$ clusters per bin. We can see that the different estimates are consistent with each other within statistical errors. The uncertainties on the sparsity of individual clusters have been evaluated through the error propagation of the cluster mass uncertainties where we have conservatively assumed a $r=0.9$ correlation, a value that is smaller than the correlation estimated in the PSZ2Lens sample\footnote{It is worth remarking that being a mass ratio, the error propagation of mass uncertainties on individual sparsity measurements result in smaller errors the larger the correlation between the mass estimates.}. In the inset plot in Fig.~\ref{fig:s200500binned} we show the average sparsity in redshift bins of size $\Delta{z}=0.1$ for the combined sample {\it Selected} LC$^2$-{\it single}+HSC-XXL, which we have used in the cosmological analysis. We have further tested the robustness of these estimates by testing the validity of the average halo sparsity consistency relations (see Appendix~\ref{app4}).

\begin{figure}[ht]
\centering
\plotone{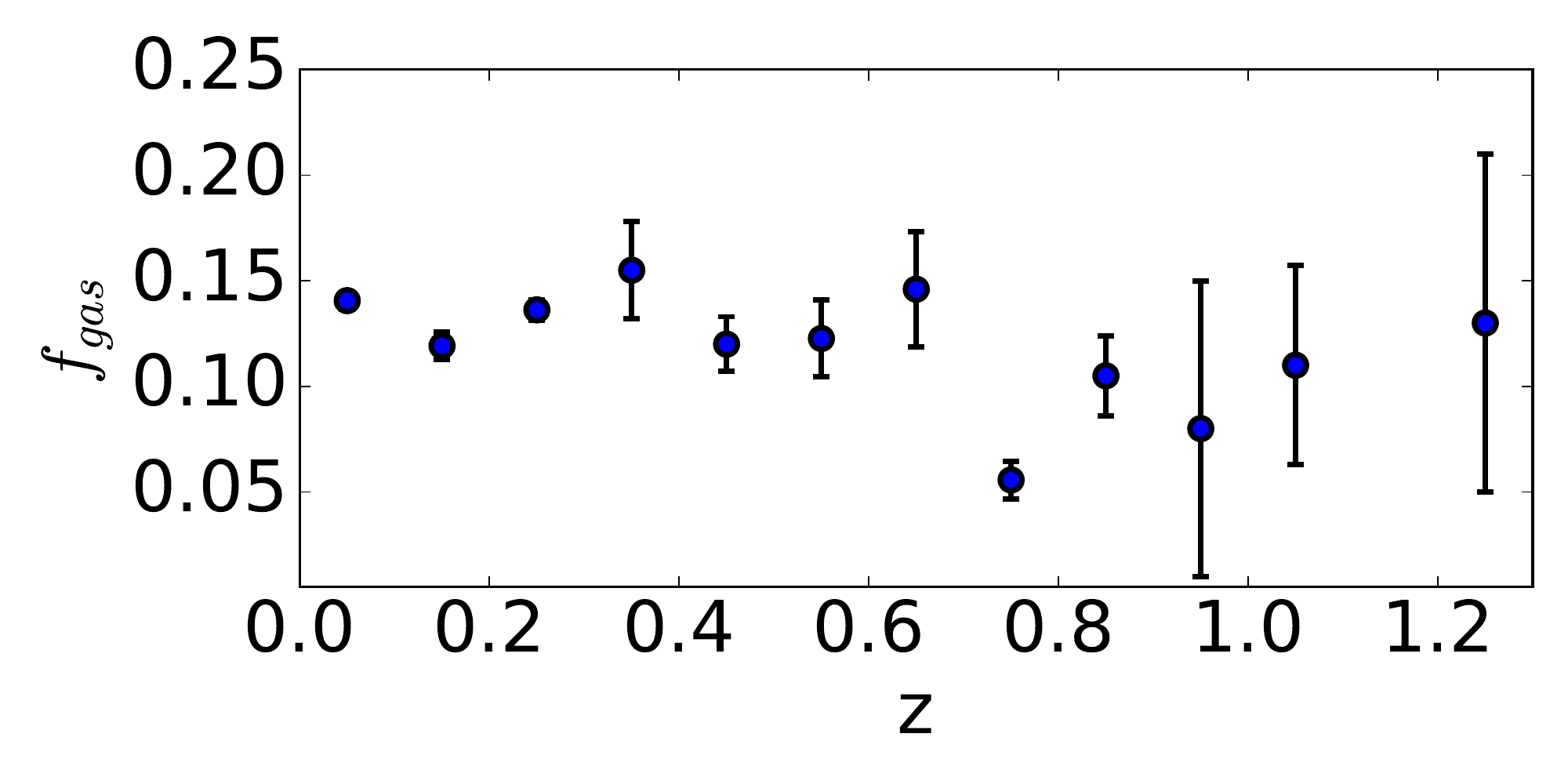} 
\caption{Averaged gas mass fraction estimates in redshift bins of size $\Delta{z}=0.1$. \label{fig:fgas}} 
\end{figure}

\subsection{Cluster Gas Mass Fraction Data}
Estimates of the gas mass fraction in clusters can be obtained from measurements of the temperature and density of the intra-cluster X-ray emitting gas. Here, we focus on measurements of $f_{\rm gas}$ within $R_{500c}$. Our dataset consists of gas mass fraction measurements of $12$ clusters at $z\lesssim 0.1$ from X-COP \citep{2019AA...621A..40E}; a sample of $44$ clusters in the range $0.1\lesssim z\lesssim 0.3$ investigated in \citet{2010A&A...524A..68E}; and a higher redshift sample of $47$ clusters from \citet{2017A&A...604A.100G} in the redshift range $0.4\lesssim z\lesssim 1.2$. The clusters from these samples are massive systems with $M_{200c}\gtrsim 10^{14}\,M_{\odot}h^{-1}$. The gas mass fraction measurements have been inferred through a backward model analysis, which fits the measured temperature profile against that predicted by solving the hydrostatic equilibrium equation \citep{2002A&A...391..841E}. In addition, we have used gas mass fraction estimates of 3 high-redshift X-ray clusters in the redshift range $0.9\lesssim z\lesssim 1.1$ derived in \citet{2018A&A...617A..64B}. The fiducial cosmology assumed in all these analyses is a flat $\Lambda$CDM model with $\Omega_m=0.3$ and $h=0.7$. We bin these measurements in redshift bins of size $\Delta{z}=0.1$ (see Fig.~\ref{fig:fgas}). 

The sparsity and gas mass fraction measurements cover a similar redshift range. It is worth remarking that the X-COP sample as well as the dataset from \citet{2010A&A...524A..68E} exploited XMM data, whereas \citet{2017A&A...604A.100G} is based on Chandra observations. Gas mass measurements are mainly based on surface brightness profiles, which are highly consistent among different X-ray observatory, whereas the dependence on temperature and metallicity is negligible.

\subsection{BAO Data}
We use cosmic distance measurements from BAO analyses to further reduce the effect of cosmological parameter degeneracies. In Table~\ref{tab_bao} we quote the measurements we have used in our analysis. These consist of estimates from the 6dF Galaxy Survey \citep[6dFGS,][]{2011MNRAS.416.3017B}, the Sloan Digital Sky Survey (SDSS) Data Release 7 Main Galaxy Sample \citep[MGS,][]{2015MNRAS.449..835R}, SDSS-III Baryon Acoustic Oscillation Spectroscopic Survey Data Release 12 \citep[BOSS-DR 12,][]{2017MNRAS.470.2617A}, SDSS-IV extended Baryon Acoustic Oscillation Spectroscopic Survey Data Release 16 Quasar Sample \citep[eBOSS-QSO,][]{2020MNRAS.499..210N}, which span a similar redshift interval as that of cluster sparsity and gas mass fraction measurements.

\begin{table}\label{tab_bao}
\centering
\begin{tabular}{ccc}
\hline\hline
Survey & $z$ & $D_V/r_d$ \\
\hline
6dFGS & $0.106$ & $2.976\pm 0.133$ \\
MGS & $0.15$ & $4.466\pm 0.168$ \\
BOSS-DR 12 & $0.38$ & $9.994\pm 0.108$ \\
BOSS-DR 12 & $0.51$ & $12.701\pm 0.128$ \\
BOSS-DR 12 & $0.61$ & $14.481\pm 0.149$ \\
eBOSS-QSO & $1.480$ & $26.51\pm 0.42$ \\
\hline
\end{tabular}
\caption{Spherically averaged cosmic distance measurements from the 6dF Galaxy Survey \citep[6dFGS,][]{2011MNRAS.416.3017B}, the Sloan Digital Sky Survey (SDSS) Data Release 7 Main Galaxy Sample \citep[MGS,][]{2015MNRAS.449..835R}, SDSS-III Baryon Acoustic Oscillation Spectroscopic Survey Data Release 12 \citep[BOSS-DR 12,][]{2017MNRAS.470.2617A}, SDSS-IV extended Baryon Acoustic Oscillation Spectroscopic Survey Data Release 16 Quasar Sample \citep[eBOSS-QSO,][]{2020MNRAS.499..210N}.}
\end{table}

\section{Cosmological Parameter Inference}\label{results}
A first attempt to perform a cosmological analysis with measurements of the internal structures of halos (using the concentration-mass relation) in combination with the gas mass fraction data has been presented in \citet{2010A&A...524A..68E}. As already mentioned, the use of the halo concentration as cosmological proxy presents several pitfalls compared to the halo sparsity. Furthermore, here we take advantage of the availability of a larger sample of cluster mass fraction data which cover a wider range of redshifts, thus more sensitive to the dependence on $D_A$, as well as the latest cosmic distance estimates from BAO analyses.

\subsection{Markov Chains \& Priors}\label{priors}
We perform a likelihood Markov Chain Monte Carlo (MCMC) data analysis of the redshift distribution of average cluster sparsity estimates in combination with BAO and gas mass fraction data to infer constraints on flat $\Lambda$CDM model specified by the following set of parameters: $\Omega_m$, $\sigma_8$, and $h$. In order to reduce the impact of parameter degeneracies, we assume a set of Gaussian priors on the remaining cosmological parameters. More specifically, we adopt a Big-Bang Nucleosynthesis \citep{2016RvMP...88a5004C} prior on the baryon density $\Omega_b h^2=0.022\pm 0.002$ and a {\it Planck}-prior \citep{2020A&A...641A...6P} on the scalar spectral index $n_s=0.965\pm 0.004$. 

We sample the target parameter space assuming uniform priors on $\Omega_m\sim U(0.1,0.9)$, $\sigma_8\sim U(0.1,1.8)$ and $h\sim U(0.55,1.20)$. Where specified we quote results obtained under a HST-prior, $h=0.7324\pm 0.0174$ \citep{2016ApJ...826...56R}, and a Planck-prior, $h=0.674\pm 0.005$ \citep{2020A&A...641A...6P}.

In the case of $f_{\rm gas}$, we assume priors on the nuisance parameters $K_{500c}$, $Y_{b,500c}$ and $f_{*,500c}$ within $R_{500c}$ (consistently with the definition of the gas mass fraction data) as determined from various works in the literature. We test the results of the parameter inference assuming different prior on $K_{500c}$. In particular, we consider the following Gaussian priors: $K^{\rm CLASH}_{500c}= 0.78\pm 0.09$ consistent with the estimate from the analysis of the CLASH sample \citep{2015MNRAS.450.3633S}; $K^{\rm CCCP}_{500c}=0.84 \pm 0.04$ as given by the analysis of a sample of clusters in \citet{2020MNRAS.497.4684H} from the Canadian Cluster Comparison Project \citep{2015MNRAS.449..685H} (CCCP); $K^{\rm CMB}_{500c}=0.65\pm 0.04$ consistent with the mass bias inferred from the joint analysis of the {\it Planck} primary CMB, the {\it Planck}-SZ number counts, the {\it Planck}-thermal SZ power spectrum and BAO in \citet{2018AA...614A..13S}. We also consider the extreme case with hard prior $K_{500c}=1$. As far as the baryon depletion factor is concerned, we assume a Gaussian prior $Y^{\rm The 300}_{b,500c}=0.938\pm 0.041$ consistently with the analysis of simulated clusters in \citet{2019AA...621A..40E} from The Three Hundred Project \citep{2018MNRAS.480.2898C}. In order to evaluate the impact of a redshift variation of the baryon depletion factor, we test the case of a Gaussian prior with mean $Y^{\rm FABLE}_{b,500c}(z)=0.931(1+0.017\cdot z+0.003\cdot z^2)$ and scatter $\sigma_{Y_{b,500c}}=0.04$ consistent with the results from the FABLE simulations \citep{2020MNRAS.498.2114H}. Finally, we assume a Gaussian prior on the stellar fraction $f_{*,500c}=0.015\pm 0.005$ consistent with estimates from a sample of clusters with masses in the same range of those of the gas mass fraction dataset \citep{2019AA...621A..40E}. Nevertheless, being a subdominant term in Eq.~(\ref{fgas}), our results are largely independent of such a prior.

\subsection{Results}
We use the Metropolis-Hastings algorithm to generate 15 independent random chains of $3\times 10^5$ samples each. We evaluate the rejection rate every 100 steps and adjust the width of the random step of the parameters dynamically. We check the convergence of the chains using the Gelman-Rubin test \citep{GelmanRubin1992} with a threshold value of $R<1.1$. We derive marginal constraints on $\Omega_m$, $\sigma_8$, $h$ and $S_8=\sigma_8\sqrt{\Omega_m/0.3}$ from the analysis of the MCMC chains. Summary tables are presented in Appendix~\ref{app5}.

\subsubsection{Sparsity \& BAO}
\begin{figure}[t]
\centering
\epsscale{1.15}
\plotone{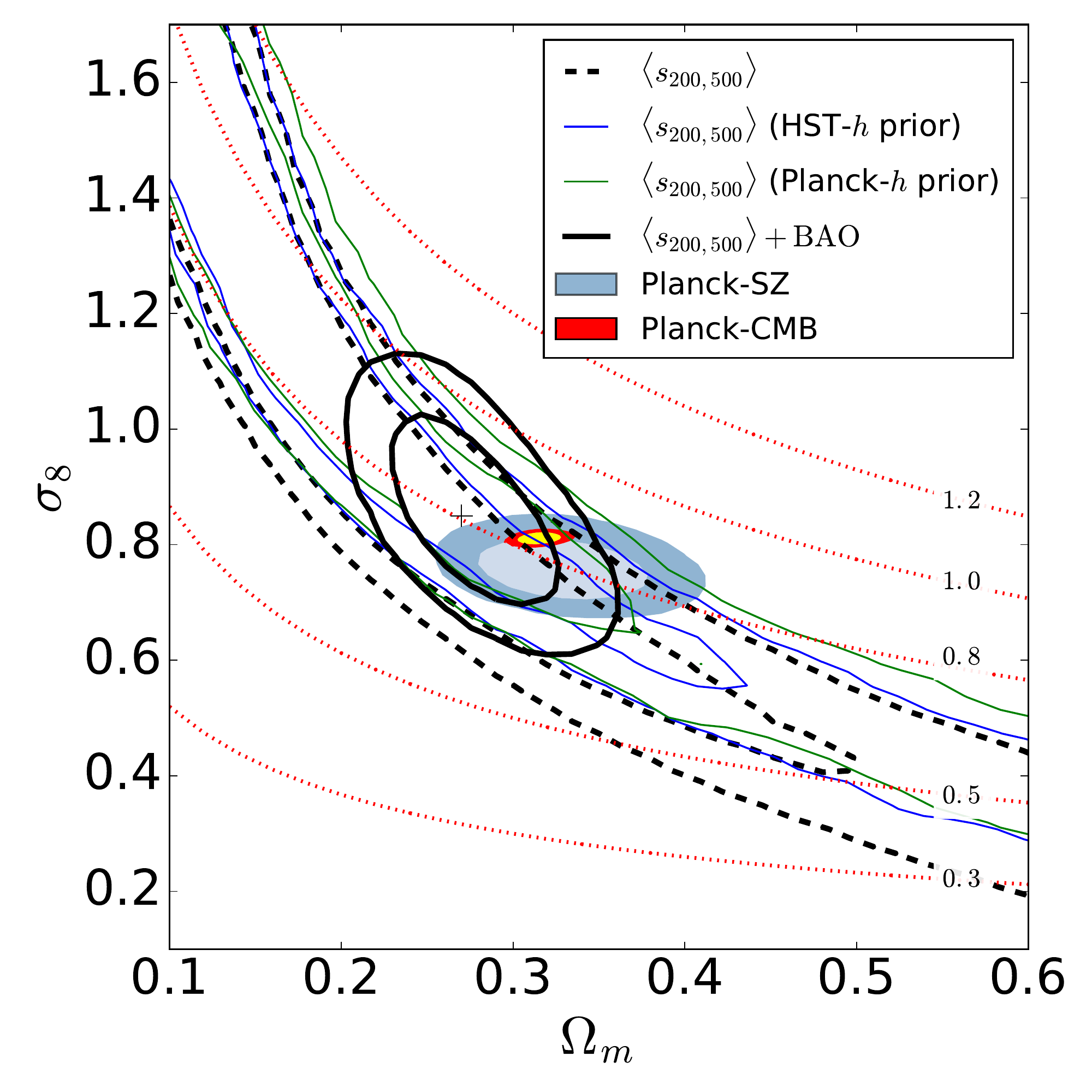}\label{fig:2dcont_spars_only_bao} 
\caption{Marginalised $1$ and $2\sigma$ contours in the $\Omega_m-\sigma_8$ from the analysis of the average sparsity data only for the uniform-$h$ (dashed black lines), HST-$h$ (blue thin lines) and Planck-$h$ (green thin lines) respectively, and in combination with BAO cosmic distance measurements (solid black lines). The cross-point corresponds to the best-fit $\Lambda$CDM model with parameter values $\hat{\Omega}_m=0.27$ and $\hat{\sigma}_8=0.85$ (and $\hat{h}=0.66$) respectively. The red dotted lines corresponds to constant $S_8$ value curves. For illustrative purposes we show the marginalised contours from the cosmological analysis of the primary {\it Planck} analysis of the CMB anisotropy power spectra \citep[TT,TE,EE+lowE+lensing,][]{2020A&A...641A...6P} and the {\it Planck}-SZ cluster analysis for their baseline model with BBN prior, BAO data and CCCP prior on the mass bias \citep[][]{2016AA...594A..24P}.} 
\end{figure}

In Fig.~\ref{fig:2dcont_spars_only_bao}, we plot the $1$ and $2\sigma$ contours in the $\Omega_m-\sigma_8$ plane from the analysis of the average sparsity measurements alone with a uniform-$h$ prior (black short-dash lines) and under the HST-prior (blue thin solid lines) and the Planck-prior (green thin solid lines) respectively. We also show the constraints derived from the combined analysis of the average sparsity with the BAO (solid black lines). We can clearly see that the constraints from the average sparsity data only are highly degenerate along curves of approximate constant $S_8$ values (red dotted lines).

In Table~\ref{table_spars_only}, we quote the mean and standard deviation of $S_8$ for the {\it Selected} LC$^2$-{\it single} clusters under different $h$-priors as well as the results obtained for the PSZ2Lens subsample. In the former case find $S_8=0.75\pm 0.20$ for the uniform-$h$ prior, while assuming the HST-prior (Planck-prior) gives $S_8=0.80\pm 0.18$ ($S_8=0.82\pm 0.16$). The mean value of $S_8$ increases for decreasing values $h$ associated to the HST and Planck priors respectively. The average sparsity also carries information on the Hubble parameter. This can be seen in Fig.~\ref{fig:2dcont_spars_only_bao}, where the contours inferred from the analysis of the sparsity data alone do shifts toward larger $S_8$ values for the HST-$h$ and Planck-$h$ priors respectively.
As it can be deduced from the values quoted in Table~\ref{table_spars_only}, the analysis of the average sparsity from the PSZ2Lens subsample gives results (bottom rows) for the different $h$-priors which are largely consistent with those from the full cluster catalog, though with slightly larger $1\sigma$ errors (of $\sim 10\%$ order) due to the smaller size and reduced redshift interval of the subsample.

Overall, the marginal constraints on $S_8$ from the analysis of the cluster sparsity only are consistent with those inferred from the sparsity analysis of a sample of X-ray clusters presented in \citet{2018ApJ...862...40C}, where it was found $S_8=0.87\pm 0.20$ for the HST-$h$ prior.

\begin{figure}[t]
\centering
\epsscale{1.05}
\plotone{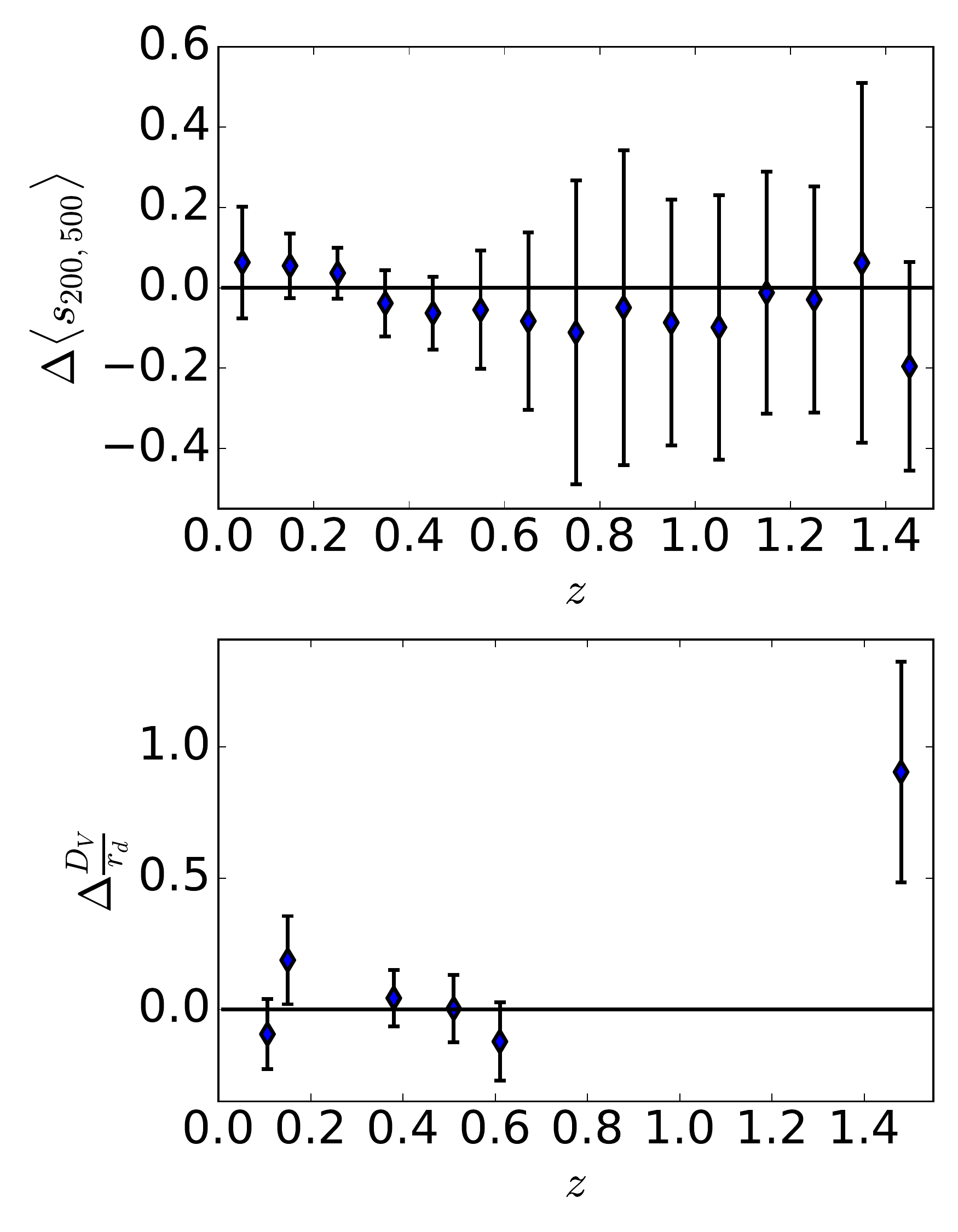}\label{spars_bao_resi} 
\caption{Residuals of the average sparsity (top panel) and BAO (bottom panel) data with respect to the best-fit $\Lambda$CDM model with parameter values $\hat{\Omega}_m=0.274$, $\hat{\sigma}_8=0.850$ and $\hat{h}=0.662$ respectively.} 
\end{figure}

Including the BAO data\footnote{We infer the following cosmological parameter constraints from the MCMC likelihood analysis of the BAO data alone for a flat $\Lambda$CDM model: $\Omega_m=0.387\pm 0.089$ and $h=0.718\pm 0.043$} allows to break the degeneracy along the $\Omega_m$ direction and infer close bounds on $\Omega_m$ and $\sigma_8$. Moreover, because of the dependence of the average sparsity on the Hubble parameter, we expect the joint analysis with the BAO data to provide closed bounds on $h$.
In order to have an appreciation of the derived constraints with respect to those inferred from other probes, in Fig.~\ref{fig:2dcont_spars_only_bao} we plot the $1$ and $2\sigma$ marginalised contours from the {\it Planck} primary CMB cosmological analysis (red-yellow filled contours) and the {\it Planck}-SZ cluster counts (blue-light blue filled contours). In Table \ref{table_spars_bao}, we quote the results of the marginal statistics from the joint sparsity and BAO analysis obtained under different $h$-priors.

We find the following marginal constraints (mean and standard deviation): $\Omega_m=0.277\pm 0.029$, $\sigma_8=0.856\pm 0.100$ (corresponding to $S_8=0.818\pm 0.070$) and $h=0.662\pm 0.023$. The marginalised posteriors are well approximated by Gaussian distributions, consequently we find the best-fit model parameters to be close to the parameter average values. In particular, we have $\hat{\Omega}_m=0.274$, $\hat{\sigma}_8=0.850$ (corresponding to $\hat{S}_8=0.813$) and $\hat{h}=0.662$. The residuals are shown in Fig.~\ref{spars_bao_resi}.

The joint analysis results in an improvement of the cosmological constraints with respect to those obtained from the BAO data alone of a factor of $3$ on $\Omega_m$ and a factor of $\sim 2$ on $h$. On the other hand, imposing the HST-$h$ (Planck-$h$) prior does not significantly improve the uncertainties on $\Omega_m$ and $\sigma_8$.

These results are consistent to better than $1\sigma$ with those inferred from the {\it Planck} primary CMB cosmological analysis of the anisotropy power spectra (red-yellow shaded contours): $\Omega_{m}=0.315\pm 0.007$, $\sigma_{8}=0.811\pm 0.006$, $S_{8}=0.832\pm 0.013$ and $h=0.6736\pm 0.0054$ \citep[TT,TE,EE+lowE+lensing, see Table 2 in][]{2020A&A...641A...6P}.

The constraints are also consistent with those of the {\it Planck}-SZ cluster counts baseline model\footnote{We are grateful to Richard Betty for providing us with the MCMC chains of the CCCP+BAO+BBN baseline model from the {\it Planck}-SZ data analysis \citep{2016AA...594A..24P}.} with $\Omega_m=0.330\pm 0.030$, $\sigma_8=0.760\pm 0.033$ and $S_8=0.796\pm 0.042$, that were derived under similar assumptions (most notably the combination of the BBN prior and BAO data). 

The comparison indicates that {\it Planck}-SZ cluster counts provides tighter constraints on $\sigma_8$, while the bounds on $\Omega_m$ have the same level of statistical uncertainty. Limits on $h$ were derived in \citet{2016AA...594A..24P}, though the values of the marginal statistics were not quoted. Here, we have analysed the chains that were made available to us and obtained $h=0.693\pm 0.027$, which is consistent with the constraints on $h$ we have inferred from the joint sparsity and BAO analysis. Nonetheless, it is worth noticing that the constraints from the {\it Planck}-SZ analysis are dominated by the BAO dataset used in \citet{2016AA...594A..24P}, as no constraints would be inferred from the use of the cluster counts only (J.~B. Melin, private communication). This is not the case of the BAO data used here.

The comparison with other cluster number counts analyses is less straightforward, since constraints on the cosmological parameters were derived under very different priors.

\begin{figure}[t]
\centering
\epsscale{1.1}
\plotone{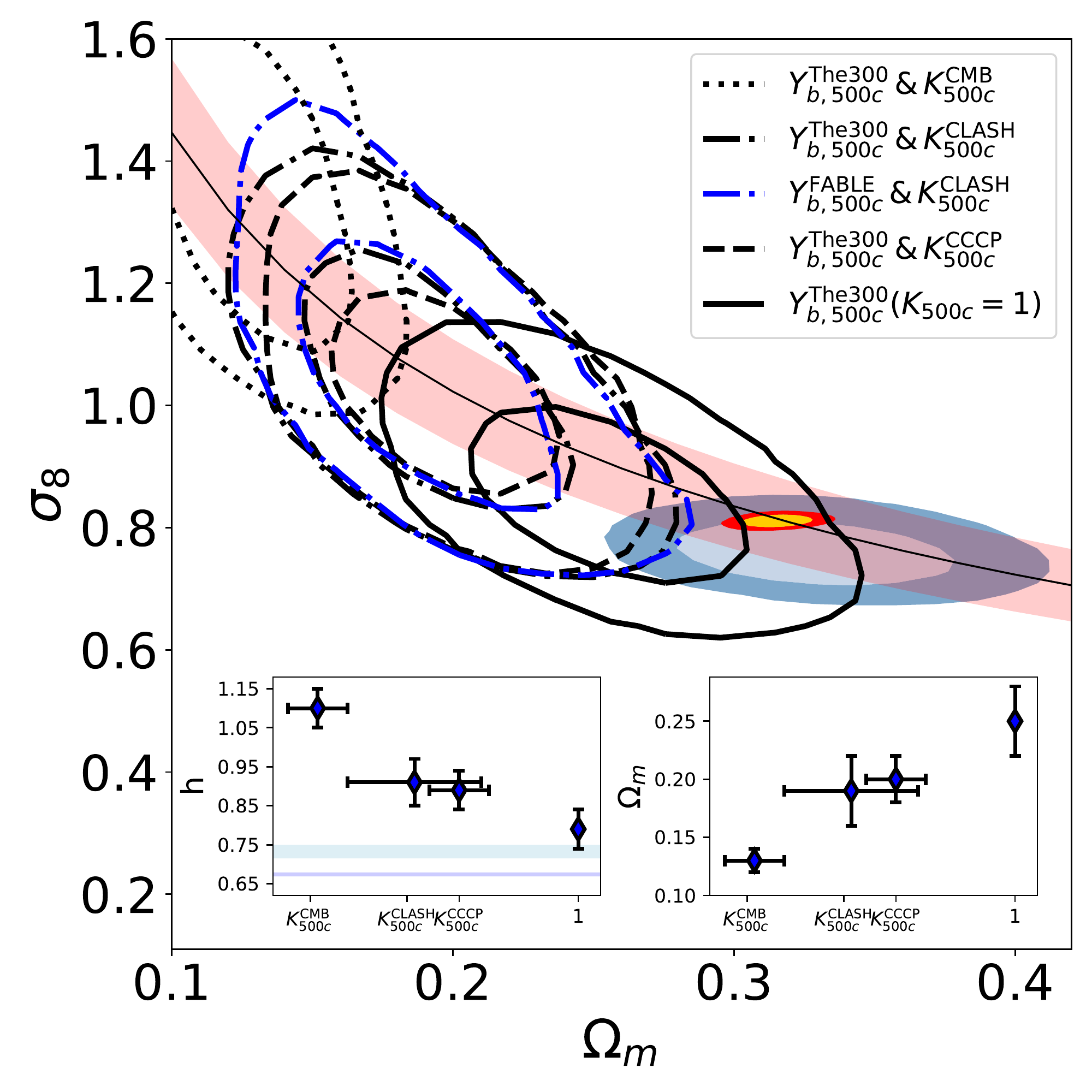}\label{fig:like2d_spars_fgas} 
\caption{Marginalised $1$ and $2\sigma$ contours in the $\Omega_m-\sigma_8$ from the combined analysis of the average sparsity and gas mass fraction data for different sets of priors: $Y_{b,500c}^{\rm The300}$ and $K^{\rm CMB}_{500c}$ (dotted black lines), $Y_{b,500c}^{\rm The300}$ and $K^{\rm CLASH}_{500c}$ (dash-dotted black lines), $Y_{b,500c}^{\rm FABLE}$ and $K^{\rm CLASH}_{500c}$ (dash-dotted blue lines), $Y_{b,500c}^{\rm The300}$ and $K^{\rm CCCP}_{500c}$ (dashed black lines), and $Y_{b,500c}^{\rm The300}$ with $K_{500c}=1$ (solid black lines). The red shaded region corresponds to curves of constant $S_8=0.83\pm 0.07$ values, that is the mean and standard deviation of the different $S_8$ estimates quoted in Table~\ref{tableparams}. As in Fig.~\ref{fig:2dcont_spars_only_bao}, we plot the contours from the {\it Planck} primary CMB and {\it Planck}-SZ cluster counts analyses respectively. The inset plots show the mean and standard deviation of $h$ (left inset) and $\Omega_m$ (right inset) as function of the $K_{500c}$ prior. In the left inset the light-blue and blue shaded areas corresponds to the HST-$h$ and Planck-$h$ priors respectively.} 
\end{figure}

\subsubsection{Sparsity \& Gas Mass Fraction}
Cluster gas mass fraction data can be used to break the $\Omega_m-\sigma_8$ degeneracy that characterises the constraints from the average sparsity. However, differently from the BAO, the possibility to derive robust constraints on $\Omega_m$, $\sigma_8$ and $h$ depends on the sensitivity of the gas mass fraction data to the priors assumed on the baryon depletion factor and the mass calibration bias. This is because such parameters appear as multiplicative factors in Eq.~(\ref{fgas}), thus resulting in the $h^{3/2}$ dependence to be highly degenerate with $\Omega_b/\Omega_m$, $K_{500c}$ and $Y_{b,500c}$.

In Fig.~\ref{fig:like2d_spars_fgas} we plot the $1$ and $2\sigma$ contours in the $\Omega_m-\sigma_8$ plane from the joint analysis of the average sparsity and gas mass fraction data obtained under different $K_{500c}$ and $Y_{b,500c}$ priors respectively. The marginal constraints on $\Omega_m$, $\sigma_8$, $S_8$ and $h$ are summarised in Table~\ref{tableparams}. The red shaded area in Fig.~\ref{fig:like2d_spars_fgas} highlights the degeneracy of the average sparsity along curves of constant $S_8=0.83\pm 0.07$ values, the mean and standard deviation of the $S_8$ estimates for the various prior configurations quoted in Table~\ref{tableparams} (top rows). The inset plots shows the marginal statistics of $h$ (left inset) and $\Omega_m$ (right inset) as function of the $K_{500c}$ prior for the $Y_{b,500c}^{\rm The300}$ case respectively. We may notice that depending on the $K_{500c}$ prior, the gas mass fraction data break the $\Omega_m-\sigma_8$ degeneracy along different locations of the $S_8$ curve. This is because for a given $Y_{b,500c}$ and $\Omega_b$ prior, there is a compensation between the fitting values of $h$ and $\Omega_m$ and the prior value of $K_{500c}$. In particular, the smaller the $K_{500c}$ prior, the larger the value of $h$, while the smaller the value of $\Omega_m$ necessary to fit the same gas mass fraction data. As a result, the gas mass fraction data breaks the $S_8$ degeneracy of the cluster sparsity at larger $\sigma_8$ values for smaller $K_{500c}$ priors. This is consistent with the trend shown in Fig.~\ref{fig:like2d_spars_fgas}, where the contours shifts from the upper left to the lower right for increasing values of the $K_{500c}$ prior, while the average inferred value of $h$ ($\Omega_m$) shown in the left  (right) inset plot decreases (increases). For comparison, we have tested the $Y_{b,500c}^{\rm FABLE}$ prior for the $K_{500c}^{\rm CLASH}$ case and find no statistically significant differences with respect to the $Y_{b,500c}^{\rm The300}$ prior. As it can be seen in Fig.~\ref{fig:like2d_spars_fgas}, only the constraints derived under the hard $K_{500c}=1$ prior are marginally consistent with those from the {\it Planck} primary CMB analysis.

We find that imposing the HST and Planck priors tends to slightly shift the parameter constraints toward larger $\Omega_m$ and smaller $\sigma_8$ values (see values quoted in the central and bottom rows in Table~\ref{tableparams}). Nevertheless, the inconsistencies among the different gas mass fraction prior model parameter assumptions remain unsolved. Most notably, in the case of the $K_{500c}^{\rm CMB}$ prior, we find constraints on $\Omega_m$ and $\sigma_8$ which lie at more than $\gtrsim 2\sigma$ from the {\it Planck} results. This suggests that the low value of $K_{500c}^{\rm CMB}$ not only stands in contrast with expectation from numerical simulations of clusters and direct estimates of the mass calibration bias from galaxy cluster samples, but also with gas mass fraction measurements we have considered here.

Given the large scatter in the prior values of the baryon depletion factor and the mass calibration (and the impact that such priors have on the cosmological parameter inference), it is advisable to simply treat them as nuisance parameters and marginalise over large uniform priors. However, because of the degeneracy with $h$, this may result into weak cosmological parameter constraints. Hence, an external independent dataset such as the BAO is required to isolate the $h$ dependence of gas mass fraction data and infer tighter bounds on the other cosmological parameters. This is what we investigate next.

\subsubsection{Joint Analysis}
Here, we present the results of the joint likelihood analysis of average cluster sparsity, gas mass fraction, and BAO data. To be as conservative as possible we infer marginalised bounds over uniform priors on the gas mass fraction nuisance parameters. More specifically, we uniformly sample the gas depletion factor $Y_{b,500c}\sim U(0.7,1.0)$ and the mass calibration bias $K_{500c}\sim U(0.6,1.0)$ over interval values which include the Gaussian priors previously discussed. 

\begin{figure}[t]
\centering
\plotone{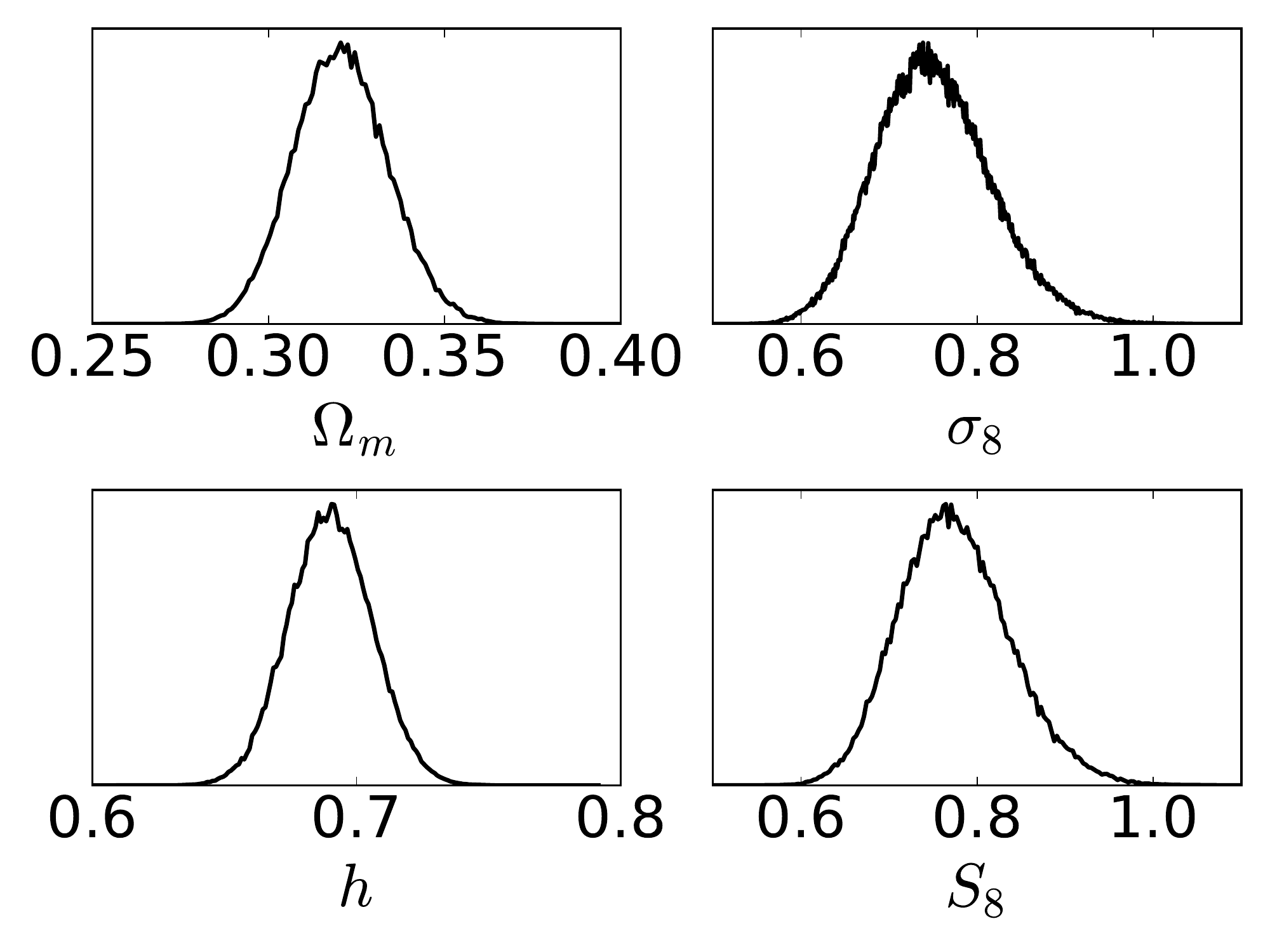}\label{fig:1djoint} 
\caption{Marginalised 1D posteriors on $\Omega_m$ (top left panel), $\sigma_8$ (top right panel), $h$ (bottom left panel) and $S_8$ (bottom right panel) from the joint analysis of cluster sparsity, gas mass fraction and BAO data.} 
\end{figure}

\begin{figure}[th]
\centering
\plotone{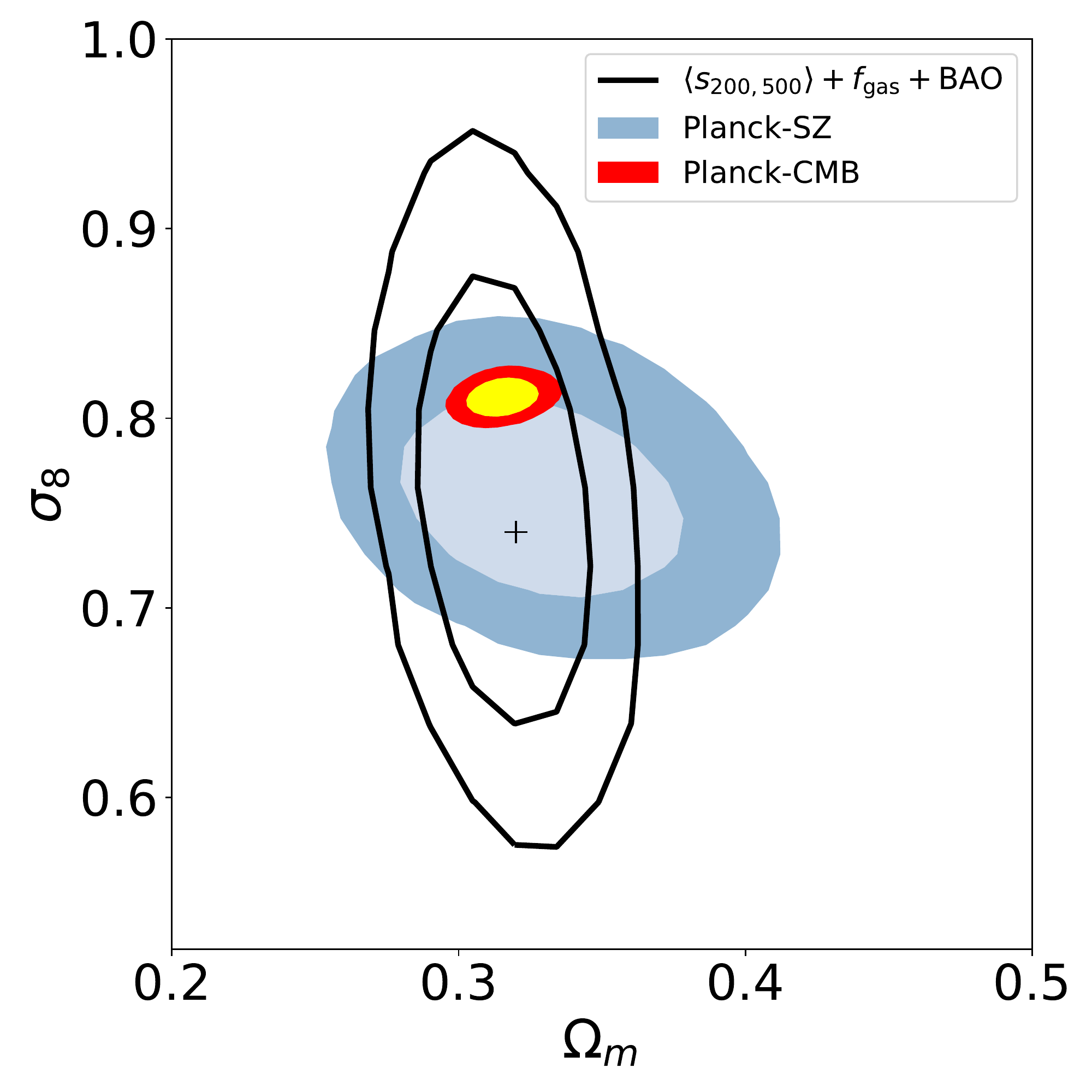}\label{fig:2dcont} 
\caption{Marginalised $1$ and $2\sigma$ contours in the $\Omega_m-\sigma_8$ from the combined analysis of the average cluster sparsity, gas mass fraction and BAO data (dot black lines). As in Fig.~\ref{fig:2dcont_spars_only_bao}, we plot marginalised contours from the {\it Planck} primary CMB analysis (yellow and red contours) and the {\it Planck}-SZ number counts (dark and light steal-blue contours). The cross-point corresponds to the best-fit $\Lambda$CDM model with parameter values $\hat{\Omega}_m=0.320$ and $\hat{\sigma}_8=0.738$ (and $\hat{h}=0.690$) respectively.} 
\end{figure}

In Fig.~\ref{fig:1djoint} we show the 1D marginalised posteriors, while in Fig.~\ref{fig:2dcont}, we plot the marginalised $1$ and $2\sigma$ contours in the $\Omega_m$-$\sigma_8$ plane. As in Fig.~\ref{fig:2dcont_spars_only_bao}, we plot the contours from the {\it Planck} primary CMB and the {\it Planck}-SZ cluster counts cosmological analyses respectively. In Table~\ref{sparsfgasbao_tab}, we quote the results of the marginal statistics of $\Omega_m$, $\sigma_8$, $S_8$ and $h$.

\begin{figure}[t]
\centering
\epsscale{1.05}
\plotone{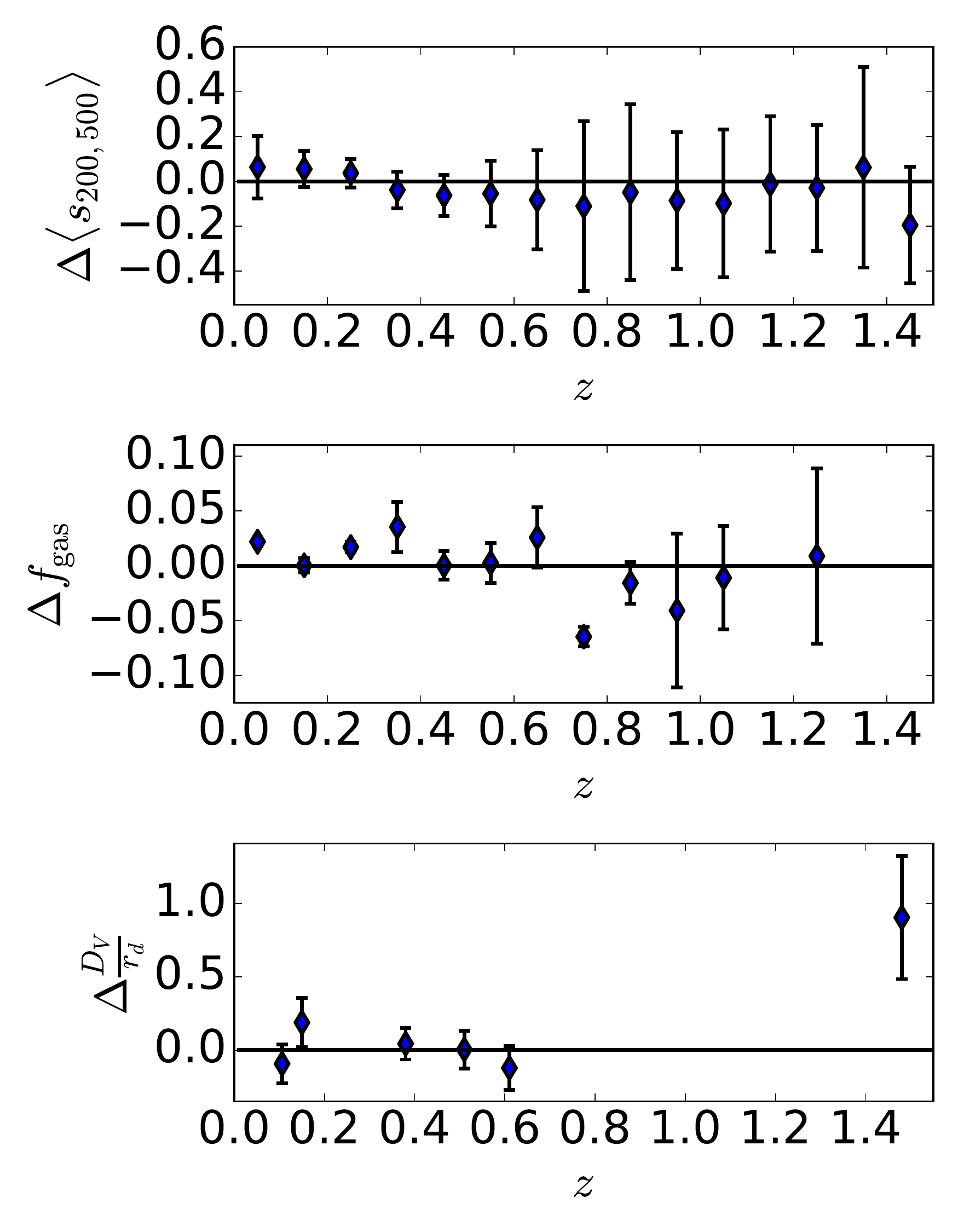}\label{sparsfgasbao_res} 
\caption{Residuals of the average sparsity (top panel), gas mass fraction (central panel) and BAO (bottom panel) data with respect to the best-fit flat $\Lambda$CDM model with parameter values $\hat{\Omega}_m=0.320$, $\hat{\sigma}_8=0.738$ and $\hat{h}=0.690$ respectively.} 
\end{figure}

We have $\Omega_m=0.316\pm 0.013$ and $\sigma_8=0.757\pm 0.067$, which corresponds to $S_8=0.776\pm 0.064$, and $h=0.6958\pm 0.0167$. We find the best-fit model parameters to be $\hat{\Omega}_m=0.320$, $\hat{\sigma}_8=0.738$ (corresponding to $\hat{S}_8=0.763$) and $\hat{h}=0.690$. The data residuals with respect to the best-fit model are shown in Fig.~\ref{sparsfgasbao_res}. Notice that the 1D marginalised posterior of $\sigma_8$ (and consequently $S_8$) has a slightly fatter tail than that of $\Omega_m$ and $h$, which are well approximated by Gaussian distributions. This is the reason of the small difference between the inferred average value of $\sigma_8$ (and $S_8$) and the best-fit value of $\hat{\sigma}_8$ (and $\hat{S}_8$) associated to the peak of the marginalised posterior. Compared to the constraints inferred from the joint analysis of the average sparsity with BAO, the addition of the gas mass fraction data improves the constraints on $\Omega_m$ by a factor $2$. The constraints on $\sigma_8$ and $h$ are improved with a gain on the $1\sigma$ errors at the $\sim 49\%$ and $\sim 35\%$ level respectively. Quite remarkably, we infer $1\sigma$ constraints on $h$ that are competitive with those from the HST analysis \citep{2016ApJ...826...56R}. As we can see in Fig.~\ref{sparsfgasbao_res}, the results of the joint analysis are consistent within $1\sigma$ with both the {\it Planck} primary CMB and the {\it Planck}-SZ cluster count analyses. Adding priors on $h$ does not significantly improve the determination of $\Omega_m$ and $\sigma_8$.

As expected we find the gas mass fraction nuisance parameters to be unconstrained in the prior parameter interval, due to the fact that they are perfectly degenerate. Hence, assuming a Gaussian prior $K_{500c}^{\rm CLASH}$ ($K_{500c}^{\rm CCCP}$) does not noticeably improve the constraints on the cosmological parameters. In contrast, it allows to infer a lower limit on $Y_{b,500c}\gtrsim 0.89$ ($Y_{b,500c}\gtrsim 0.91$) at $1\sigma$. 

\begin{figure}[t]
\centering
\plotone{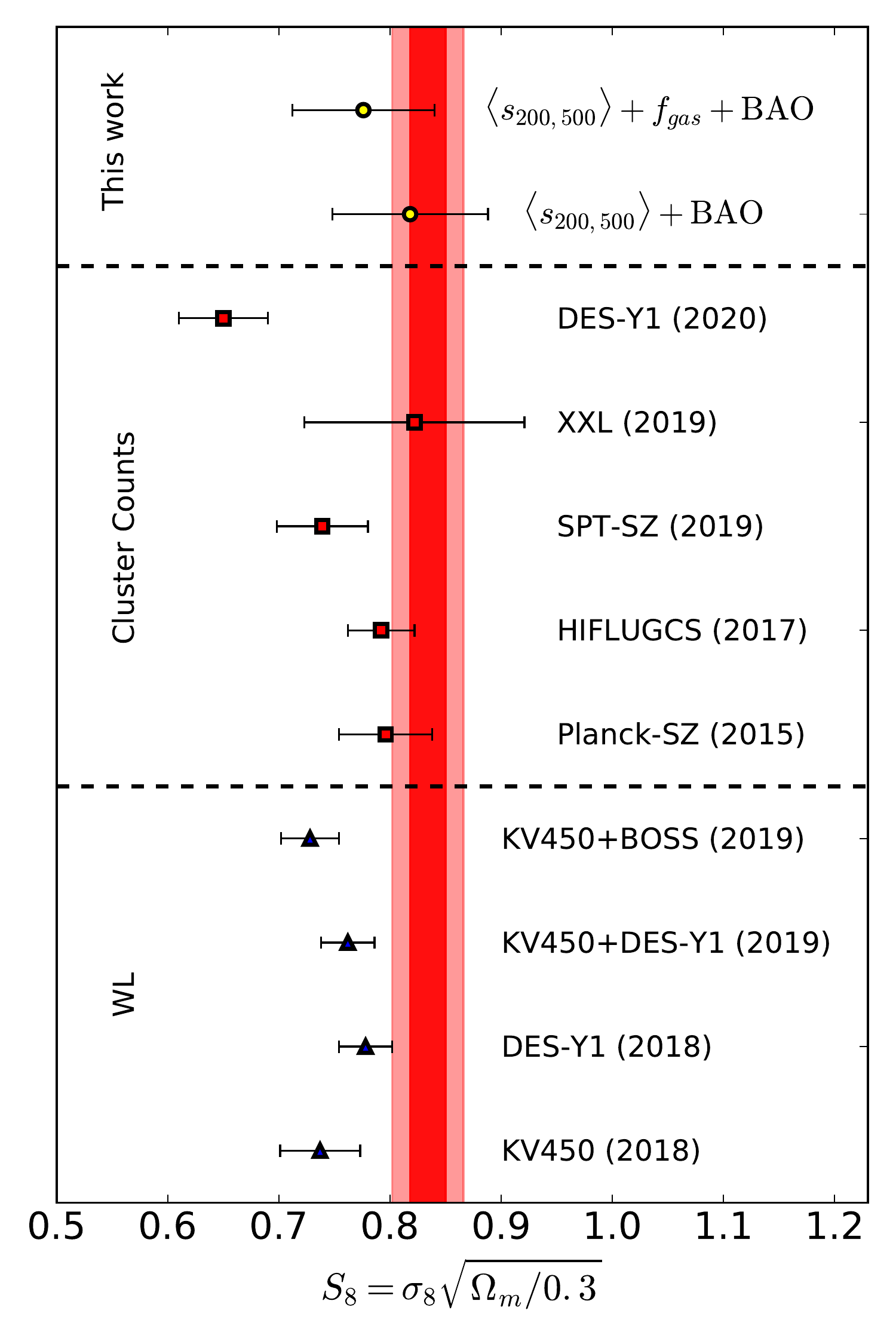}\label{fig:S8}
\caption{Constraints on $S_8$ from the analysis of shear lensing data (blue triangles) of KV450 \citep{2020A&A...633A..69H}, DES-Y1 \citep{2018PhRvD..98d3528T}, KV450+DES-Y1 \citep{2020A&A...638L...1J}, KV450+BOSS \citep{2020A&A...633L..10T}; cluster number counts (red squares) of {\it Planck}-SZ \citep{2016AA...594A..24P}, HIFLUGCS \citep{2017MNRAS.471.1370S}, SPT-SZ \citep{2019ApJ...878...55B}, XXL \citep{2018AA...620A..10P} and DES-Y1 \citep{2020PhRvD.102b3509A}.  The constraints inferred from the combined analysis of cluster sparsity and BAO (upper yellow circle) and together with the gas mass fraction (lower yellow circle) data are shown at the top of the plot. The shaded area corresponds to the $1$ and $2\sigma$ limits from the {\it Planck} primary CMB analysis \citep[TT,TE,EE+lowE+lensing,][]{2020A&A...641A...6P}.} 
\end{figure}

\begin{figure}[th]
\centering
\plotone{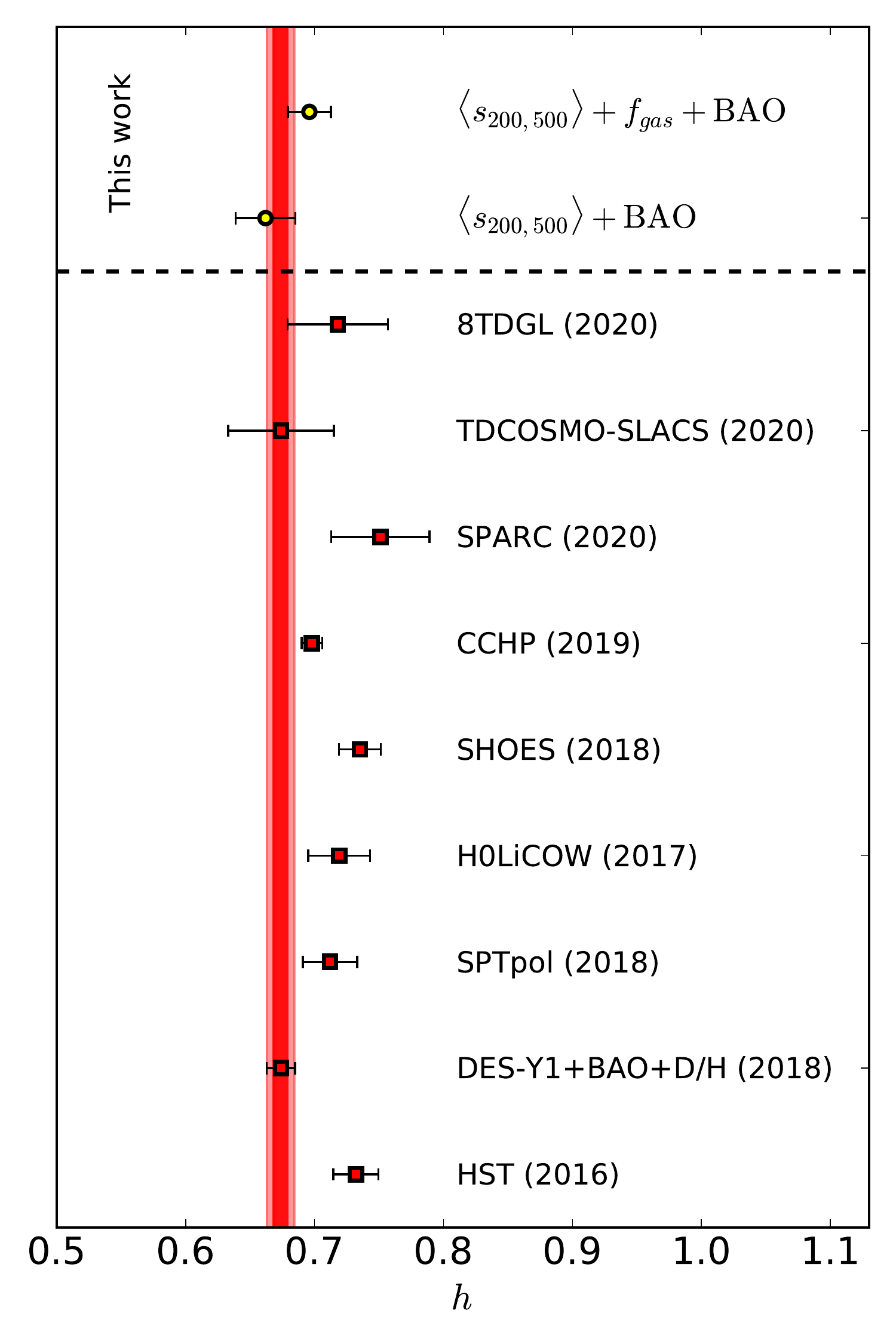}\label{fig:h}
\caption{Constraints on $h$ from various probes: HST \citep{2016ApJ...826...56R}, DES-Y1+BAO+D/H \citep{2018MNRAS.480.3879A}, SPTpol \citep{2018ApJ...852...97H}, H0LiCOW \citep{2017MNRAS.465.4914B}, SHOES \citep{2018ApJ...861..126R}, CCHP \citep{2019ApJ...882...34F}, SPARC \citep{2020AJ....160...71S}, TDCOSMO-SLACS \citep{2020arXiv200702941B}, 8TDGL \citep{2020arXiv200714398D}. The constraints inferred from the combined analysis of cluster sparsity and BAO (upper yellow circle) and together with the gas mass fraction (lower yellow circle) data are shown at the top of the plot. The shaded area corresponds to the $1$ and $2\sigma$ limits from the {\it Planck} primary CMB analysis \citep[TT,TE,EE+lowE+lensing,][]{2020A&A...641A...6P}.} 
\end{figure}

In Fig.~\ref{fig:S8} and Fig.~\ref{fig:h} we show summary plots of state-of-art estimates of $S_8$ \citep[see also][]{2019SSRv..215...25P} and $h$ from various probes.

\section{Conclusions}\label{conclusions}
In recent years, the increased availability of large, complete samples has opened the way to probing cosmology with galaxy cluster observations. Cosmological parameter constraints have been primarily inferred from cluster number count data analyses. These have provided constraints complementary to those inferred from other standard probes such as the CMB anisotropy power spectra. However, several source of systematics may still affect the results of these studies. As an example, cluster number count measurements from SZ catalogs have resulted in values of $S_8$ which are lower than those obtained from the analysis of the {\it Planck} anisotropy power spectra  \citep[see e.g.][]{2016AA...594A..24P,2019ApJ...878...55B,2020PhRvD.102b3509A}. Errors in the mass calibration of clusters as well as selection effects may be responsible for such discrepancies, although it cannot be {\it a priori} excluded the effect of novel physics not in the standard $\Lambda$CDM model. For this reason, it is timely to investigate other probes of galaxy cluster cosmology that may provide independent constraints on the cosmological parameters. Besides cluster number counts, measurements of the spatial clustering of galaxy clusters \citep{2018A&A...620A...1M} as well as their internal mass distribution carry cosmological information. As an example, \citet{2014MNRAS.437.2328B} have shown that measurements of the dark matter halo sparsity can retrieve the cosmological signal encoded in the mass profile of halos hosting galaxy clusters. Moreover, these measurements have been shown to be insensitive to selection effects and mass calibration errors \citep{2018ApJ...862...40C}.

Here, we have derived cosmological constraints from estimates of the average sparsity of galaxy clusters using lensing mass measurements. We have used lens mass measurements from a selected sample of clusters from the LC$^2$-{\it single} catalog \citep{2015MNRAS.450.3665S} together with HSC-XXL clusters \citep{2020ApJ...890..148U}. We have discussed different sources of systematic errors. In order to break cosmological parameter degeneracies we have performed a combined MCMC likelihood analysis of average sparsity estimates in combination with gas mass fraction and BAO data. We find the combination of these dataset to provide competitive constraints on $\Omega_m$, $\sigma_8$ and $h$ which are consistent with those from the {\it Planck} primary CMB and the {\it Planck}-SZ cluster counts. In particular, we find $\Omega_m=0.316\pm 0.013$, $\sigma_8=0.757\pm 0.067$ (resulting in $S_8=0.776\pm 0.064$) and $h=0.696\pm 0.017$. 

Compared to other cosmological proxies from galaxy cluster observations, the sparsity has the advantage of being less sensitive to systematic errors due to mass calibration bias and selections effects. In the future, cosmological parameter uncertainties can be further reduced thanks to the availability of larger galaxy cluster samples with improved mass estimates, as well as better gas mass fraction measurements. It is possible that the combined analysis of cluster sparsity and gas mass fraction data together with estimates of the cluster abundance may not only provide strong cosmological parameter constraints, but also allow for an accurate determination of astrophysical model parameters such as the gas depletion factor and the mass calibration bias, that are sources of systematic uncertainty associated with gas mass fraction and cluster number count data analyses. We leave this to a future investigation.

A first step in this direction will be the joint study of the sparsity and gas mass fraction in galaxy clusters from a well-selected sample of a large number ($\sim 100$) of objects homogeneously analysed in their X-ray and lensing signal out to $R_{500}$ and beyond, as the one that will be available from the CHEX-MATE project\footnote{\url{http://xmm-heritage.oas.inaf.it/}} \citep{2020arXiv201011972T}.

\acknowledgments
PSC is grateful to Iacopo Bartalucci for providing data from his X-ray cluster analysis, Richard Battye for providing the {\it Planck}-SZ chains, Jean-Baptiste Melin for discussions on the {\it Planck}-SZ analysis and Florian Pacaud for providing the values of $S_8$ obtained from the XXL cluster number counts. SE and MS acknowledge financial contribution from contract ASI-INAF n.2017-14-H.0 and INAF `Call per interventi aggiuntivi a sostegno della ricerca di main stream di INAF' n. 1.05.01.86.10. The research leading to these results has received funding from the European Research Council under the European Union's Seventh Framework Programme (FP/2007--2013) / ERC Grant Agreement n.~279954. Data visualisations are prepared with the \textsc{\mbox{matplotlib}}\footnote{\url{http://matplotlib.org/}} library \citep{Hunter07}.

\bibliographystyle{aasjournal}

\appendix

\section{Average Sparsity Redshift Model Correction}\label{app1}
\begin{figure}[t]
\centering
\epsscale{0.65}
\plotone{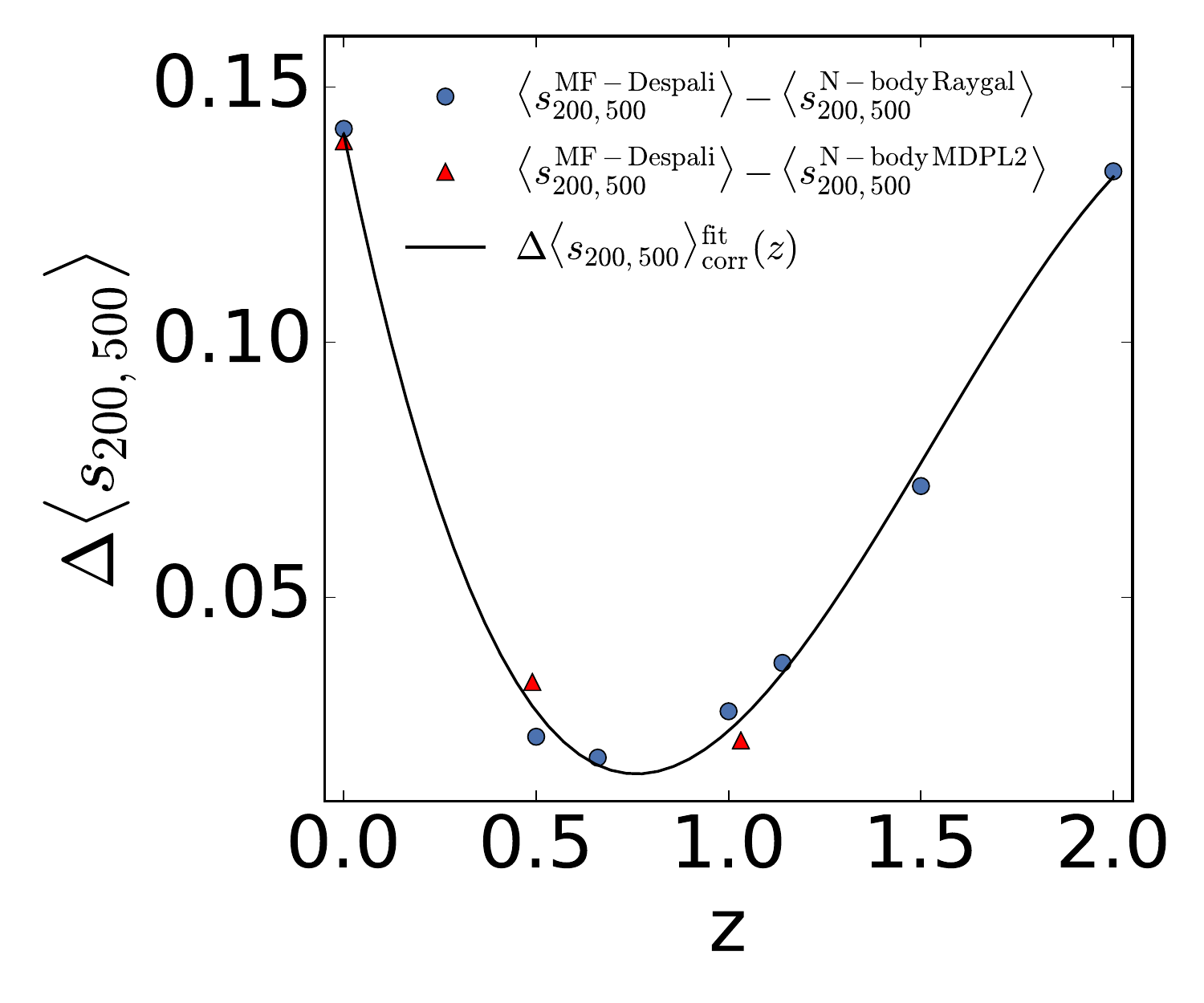}\label{fig:spars_corr}
\caption{Difference between the prediction of the average sparsity at $\Delta=200$ and $500$ using the ST-Despali parametrization and the average sparsity from the N-body halo catalogs of the RayGalGroupSims (blue dots) and MultiDark-Planck2 (red triangles) at the redshift snapshots of the simulations.} 
\end{figure}

The validity of Eq.~(\ref{spars_pred}) in predicting the ensemble average halo sparsity has been extensively tested using  N-body halo catalogs from the RayGalGroupSims simulation of a $\Lambda$CDM model in \citet[][]{2018ApJ...862...40C} and the halo catalogs from the MultiDark-Planck2 simulation in \citet{2019MNRAS.487.4382C}. These studies have shown that the simulation calibrated mass functions can predict the average halo sparsity to better then a few percent accuracy level. On the other hand, as already mentioned in Section~\ref{method}, assuming the ST-Despali parametrisation, \citet{2018ApJ...862...40C} found that the predicted average sparsity is consistent with the RayGalGroupSims results within a few percent level only in the redshift range $0.5\lesssim z\lesssim 1.2$, while differences increase up to $\sim 10\%$ at lower and higher redshifts. Such discrepancy is essentially due to the fact that the halo catalogs used in the calibration of the ST-Despali parametrisation contain different halos at different overdensities, rather than identical halos with masses estimated at different overdensities as expected from Eq.~(\ref{spars_pred}). This induces a selection effect on the calibrated mass function at the two overdensities of interest that propagates into average sparsity prediction. This can be seen in Fig.~\ref{fig:spars_corr}, where we plot the difference between the value of the average sparsity from the ST-Despali parametrisation and that measured from the N-body halos at the redshift snapshots of the RayGalGroupSims (blue dots) and the MultiDark-Planck2 (red triangles) simulations respectively. Quite importantly, we may notice that the systematic difference of the ST-Despali prediction with respect to these N-body results is similar, despite the fact that the two simulations do not share the same cosmology and do not posses similar characteristics\footnote{The MultiDark-Planck2 simulation consists of a $(1\,{\rm Gpc}\,h^{-1})^3$ volume with $3840^3$ N-body particles (corresponding to a mass resolution of $m_p=1.51\cdot 10^9\,M_{\odot}h^{-1}$) of a flat $\Lambda$CDM model with $\Omega_m=0.3071$, $\Omega_b=0.048206$, $h=0.678$, $n_s=0.96$ and $\sigma_8=0.823$ \citep{2016MNRAS.457.4340K}; the RayGalGroupSims simulation has a $(2.625\,{\rm Gpc}\,h^{-1})^3$ volume with $4096^3$ N-body particles (corresponding to a mass resolution $m_p=1.88\cdot 10^{10}\,M_{\odot}h^{-1}$) of a flat $\Lambda$CDM model with $\Omega_m=0.2573$, $\Omega_b=0.048356$, $h=0.72$, $n_s=0.963$ and $\sigma_8=0.801$.}, thus highlighting the algorithmic nature of this selection effect. As the RayGalGroupSims has a volume $\sim 8$ times larger than that of the MultiDark-Planck2 run, it allows us to precisely estimate the average sparsity at $z\gtrsim 1$ where the abundance of massive halos hosting galaxy groups and clusters drops rapidly. Henceforth, we can correct the systematic shift of the ST-Despali prediction shown in Fig.~\ref{fig:spars_corr}, by introducing a third-order polynomial function of redshift which well approximate the numerical simulation points:
\begin{equation}\label{spars_corr}
\Delta\langle s_{200,500}\rangle_{\rm corr}^{fit}(z)=s_0+s_1\cdot z+s_2\cdot z^2+s_3\cdot z^3,
\end{equation}
with $s_0=0.14073268$, $s_1=-0.37056373$, $s_2=0.32410084$ and $s_3=-0.07044901$.

\section{Biased Sparsity Estimation from Shear Lensing Profile Fits}\label{app2}
Weak lensing masses are usually inferred by fitting the shear lensing measurements against a parametric shear profile. Several of these estimate in the literature fit the data with a 1-parameter profile such as the SIS profile or the NFW \citep{1997ApJ...490..493N} with fixed concentration-mass relation. However, this results in biased estimates of the halo sparsity.

\begin{figure}[t]
\epsscale{0.75}
\centering
\plotone{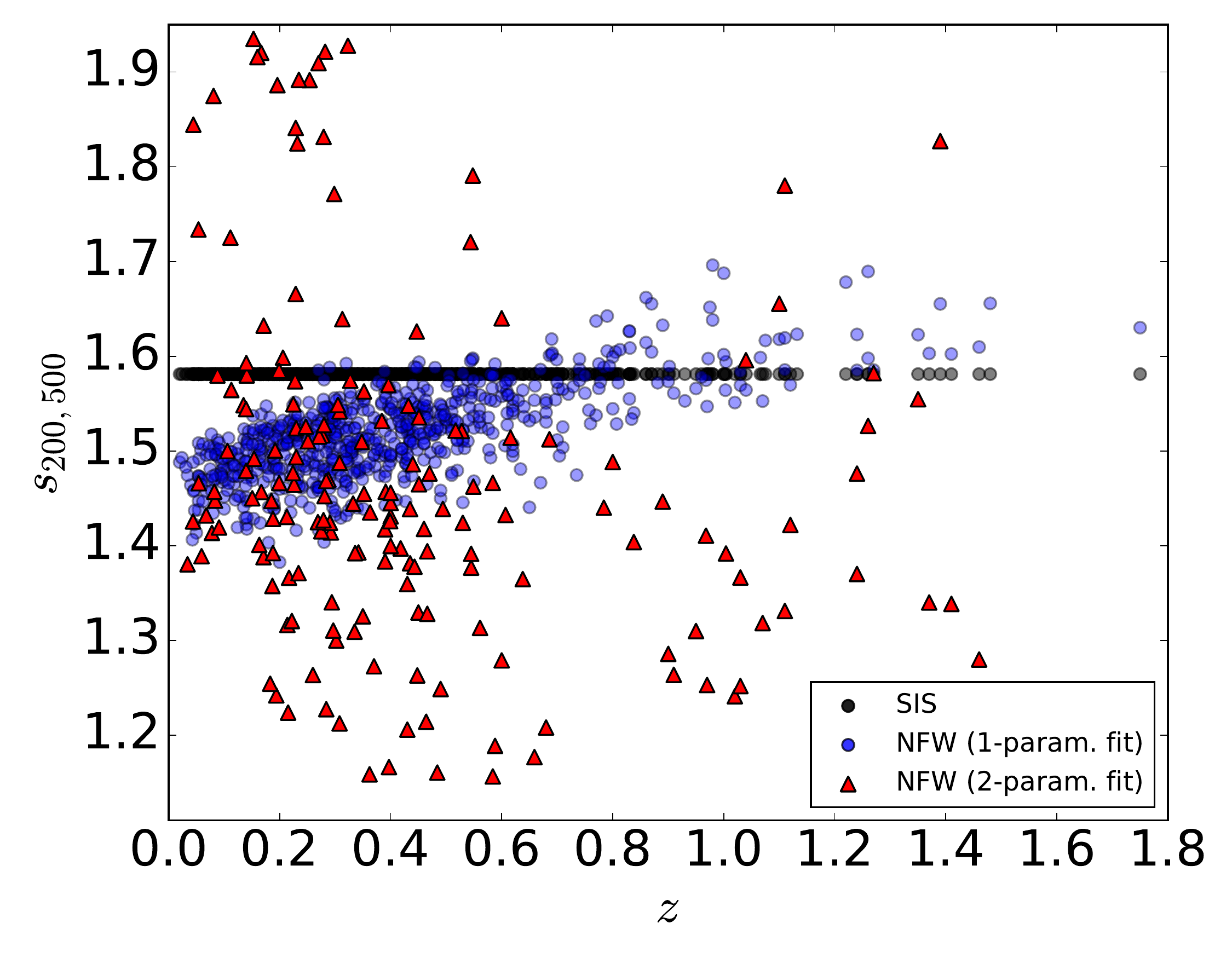}\label{fig:spars_bias}
\caption{Sparsity of clusters from mass measurements inferred fitting the SIS profile (black dots), the one-parameter fit NFW-profile with the concentration-mass relation from \citet{2008MNRAS.390L..64D} (blue dots) and a two-parameter fit NFW-profile (red triangles).} 
\end{figure}

In the case of the SIS profile this can be proven analytically, since assuming a radial density profile of the form $\rho(r)\propto 1/r^2$ gives a mass within a radius $r_{\Delta}$ enclosing an overdensity $\Delta$ that is $M_{\Delta}\propto 1/\sqrt{\Delta}$. Consequently, assuming a SIS profile results in a redshift and cosmology independent sparsity such that assuming $\Delta=200$ and $500$ gives $s^{\rm SIS}_{200,500}=\sqrt{500/200}$ contrary to the N-body simulation results. In Fig.~\ref{fig:spars_bias}, we plot the biased sparsity values for a cluster sample with masses $M_{200c}$ equal to the LC$^2$-single clusters. The SIS estimates correspond to the strake of black points distributed along the straight line at $s_{200,500}\simeq 1.58$. 

Similarly, sparsity estimates from mass measurements obtained assuming the NFW profile with a fixed concentration-mass relation, which leaves only one free parameter, cannot probe the full sparsity range either. Fixing the concentration-mass relation artificially reduces the scatter in the distribution of sparsity estimates compared to expectations from N-body simulations \citep[see e.g.][for a recent dedicated analysis]{2021MNRAS.500.5056R}. This can be seen in Fig.~\ref{fig:spars_bias} where the blue dots corresponds to sparsity estimates from mass measurements obtained from a one parameter fit NFW profile assuming the concentration-mass relation from \citet{2008MNRAS.390L..64D}. For comparison, the red triangles show the estimates from a two parameter fit NFW profiles. Hence, to avoid any bias in our analysis, we focus on lensing mass measurements inferred from two-parameter analyses. It is worth remarking that weak lensing masses are measured assuming a cosmological framework, however such dependence cancels out when taking mass ratios \citep{2015MNRAS.450.3665S}.

\section{Galaxy Cluster Sparsity Outliers}\label{app3}
\begin{figure}[t]
\epsscale{0.55}
\centering
\plotone{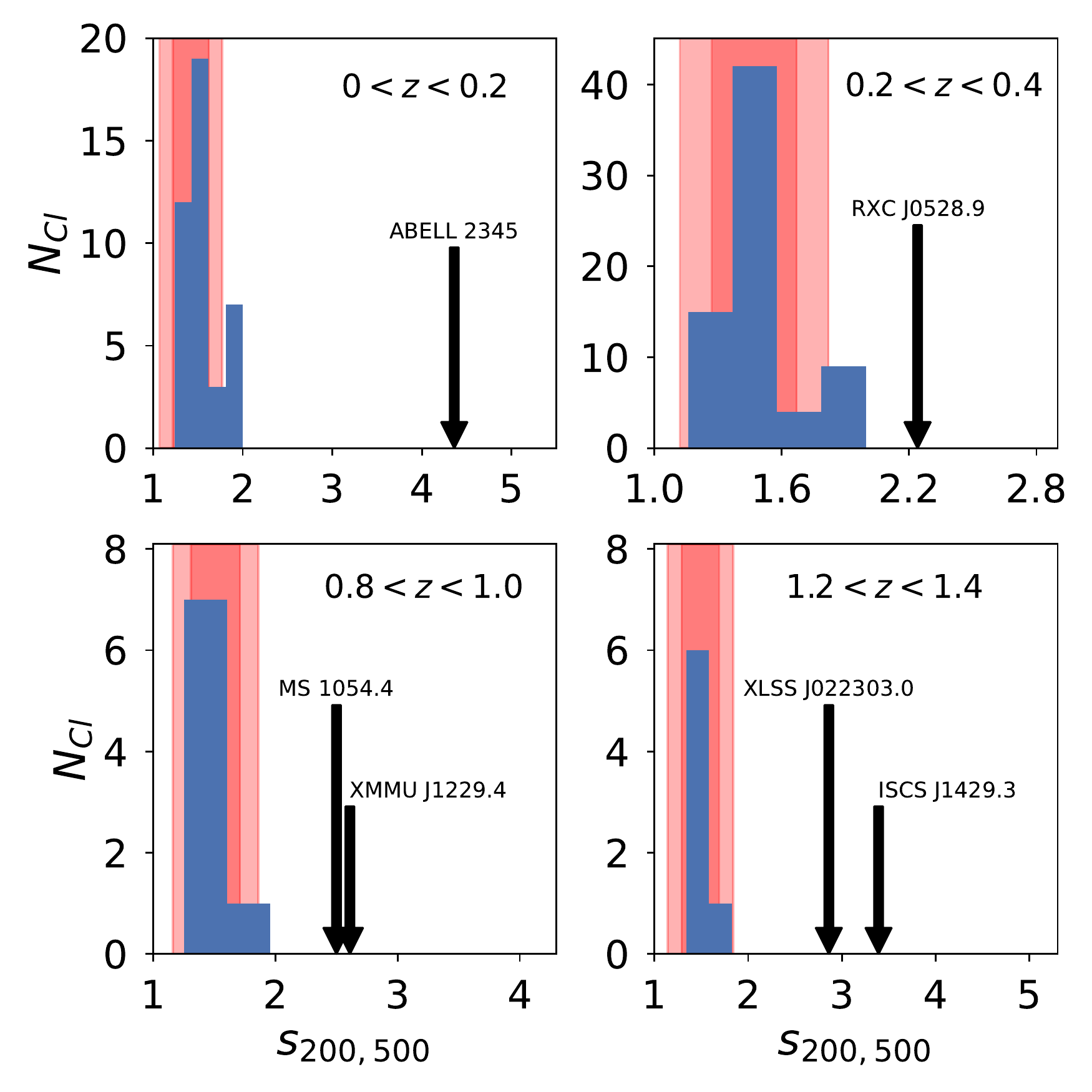}\label{fig:outliers}
\caption{Distribution of the sparsity of the {\it Selected} LC$^2$-{\it single} clusters in redshift bins of size $\Delta{z}=0.2$  containing Abell 2345 (top left panel), RXC J0528.9-3927 (top right panel), MS 1054.4-0321 and XMMU J1229.4+0151 (bottom left panel), XLSS J022303.0-043622 and ISCS J1429.3+3437 (bottom right panel). The shaded area corresponds to the $1$ and $2\sigma$ scatter of the halo sparsities estimated for the different redshift bins from the N-body halo catalogs of the RayGalGroupSims simulation.} 
\end{figure}

We have identified $6$ clusters in the LC$^2$-{\it single} sample which exhibit sparsity values that significantly differ from the distribution of the sparsity of clusters in the same redshift bin as shown in Fig.~\ref{fig:outliers}. These are Abell 2345, RXC J0528.9-3927, MS 1054.4-0321, XMMU J1229.4+0151, XLSS J022303.0-043622 and ISCS J1429.3+3437. As we can see, outliers have values of the sparsity $s_{200,500}>2.2$. Such large values are a characteristic of unrelaxed/perturbed systems. Abell 2345 is a cluster at $z=0.176$, which appears to be constituted by two merging sub-clusters \citep{2009A&A...494..429B,2010A&A...521A..78B}. Similarly, RXC J0528.9-3927 ($z=0.284$) \citep{2017A&A...601A.145F} and MS 1054.4-0321 ($z=0.83$) \citep{2000ApJ...539..540C} appear to be not fully relaxed clusters. XMMU J1229.4+0151 ($z=0.98$) may also be a perturbed system, since the lensing convergence map shows the presence of a second strong peak off-centred with respect to the position of the peak of the X-ray emission and that of the cluster galaxies \citep{2011ApJ...737...59J}. Finally, XLSS J022303.0-043622 and ISCS J1429.3+3437 are among the most distant clusters detected to date at $z=1.22$ and $1.26$ respectively. However, we found no information available in the literature on their structural and dynamical properties that might account for their large values of the sparsity. In order to compare these outliers with the expected distribution of halo sparsities from numerical simulations, we plot in Fig.~\ref{fig:outliers} the $1$ and $2\sigma$ scatter of the halo sparsity from catalogs of the RayGalGroupSims simulation presented in \citet{2018ApJ...862...40C}.

\begin{figure}
\epsscale{0.53}
\centering
\plotone{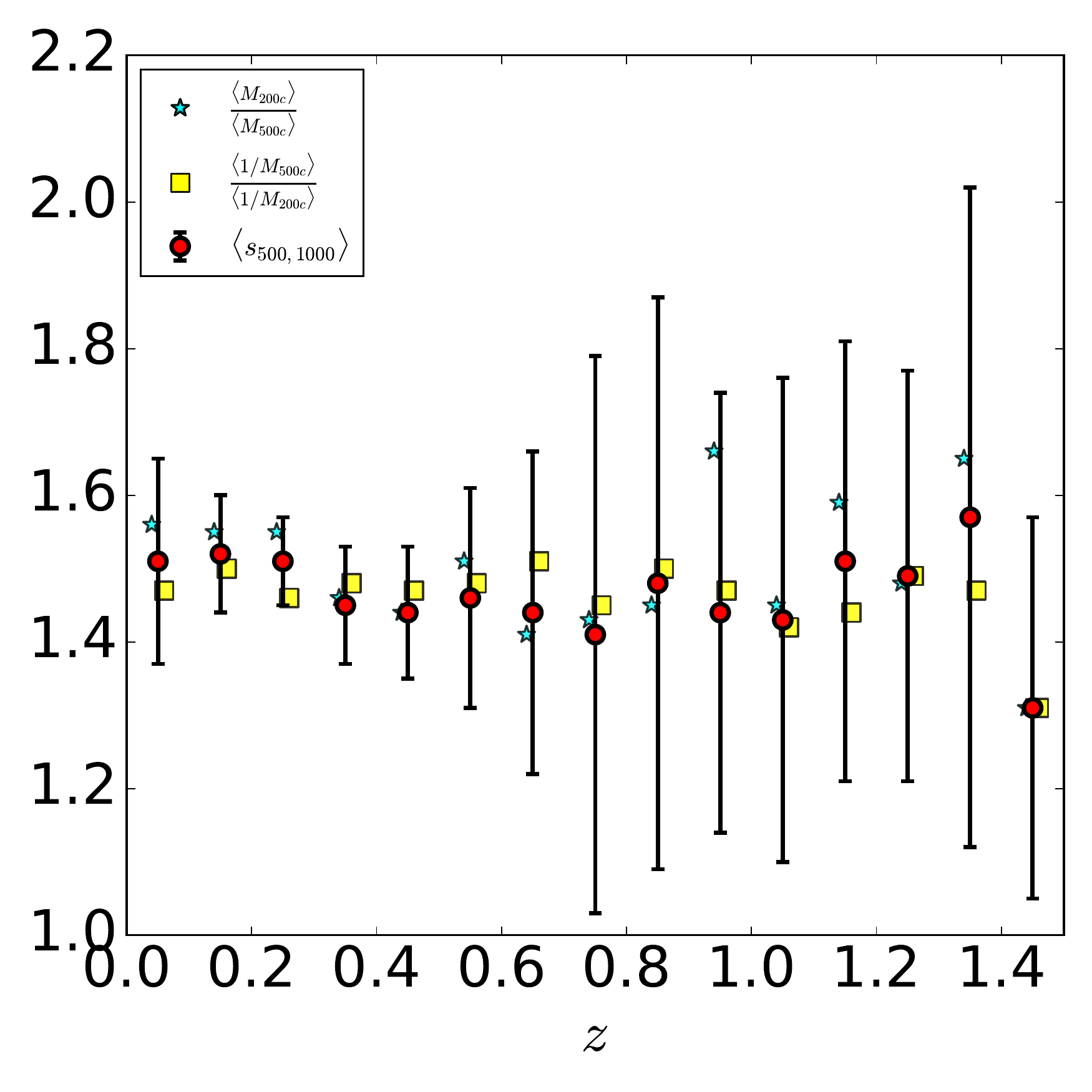}\label{fig:spars_consistency}
\caption{Ensemble average cluster sparsity for the combined sample {\it Selected} LC$^2$-{\it single} and HSC-XXL (red points) in redshift bins of size $\Delta{z}=0.1$. The estimates given by $\langle M_{200c}\rangle/\langle M_{500c}\rangle$ (cyan stars) and $\langle 1/M_{500c}\rangle/\langle 1/M_{200c}\rangle$ (yellow squares) are consistent with the ensemble average values well within the uncertainties due to mass measurement errors.} 
\end{figure}

\section{Average Sparsity Consistency Relations}\label{app4}
\citet{2019MNRAS.487.4382C} have shown that the validity of Eq.~(\ref{spars_pred}) also implies the validity of the following relations:
\begin{equation}
\langle s_{200,500}\rangle\equiv\bigg \langle\frac{M_{200c}}{M_{500c}}\bigg\rangle \approx \frac{\langle M_{200c}\rangle}{\langle M_{500c}\rangle} \approx \frac{\langle 1/M_{500c}\rangle}{\langle 1/M_{200c}\rangle}.
\end{equation}
These provide a set of consistency relations that can be used to test the consistency of sparsity measurements in galaxy cluster samples \citep{2019MNRAS.487.4382C}. Here, we compute the different average sparsity estimates for the combined {\it Selected} LC$^2$-{\it single}+HSC-XXL sample in redshift bins of size $\Delta{z}=0.1$. The results are shown in Fig.~\ref{fig:spars_consistency}. We can see that the various determinations are consistent with the ensemble average sparsity estimates within the uncertainties due to mass measurement errors.

\section{MCMC Likelihood Analysis Marginal Statistics}\label{app5}
We quote the mean and standard deviation of the different cosmological model parameters inferred from the MCMC likelihood analysis of the average sparsity data only in Table~\ref{table_spars_only}, the average sparsity in combination with BAO in Table~\ref{table_spars_bao} and in combination with the gas mass fraction in Table~\ref{tableparams}, while the results of the joint analysis are given in Table~\ref{sparsfgasbao_tab}.

\begin{table}[th]
\centering
Sparsity Only \\
\begin{tabular}{cc}
\hline\hline
Cluster Sample \& $h$-prior & $S_8$ \\
 \hline\hline
$Selected$ LC$^2-Single$ + Uniform & $0.75\pm 0.20$ \\
$Selected$ LC$^2-Single$ + HST & $0.80\pm 0.18$ \\
$Selected$ LC$^2-Single$ + Planck & $0.82\pm 0.16$ \\
\hline
PSZ2Lens + Uniform & $0.69\pm 0.22$ \\
PSZ2Lens + HST & $0.73\pm 0.21$\\
PSZ2Lens + Planck  & $0.75 \pm 0.19$\\
\hline
\end{tabular}
\caption{Mean and standard deviation of $S_8$ from the analysis of the average sparsity of the $Selected$ LC$^2-Single$ catalog (top rows) and the PSZ2Lens subsample (bottom rows).}
\label{table_spars_only}
\end{table}

\begin{table}[th]
\centering
Sparsity + BAO \\
\begin{tabular}{ccccc}
\hline\hline
 $h$-prior & $\Omega_m$ & $\sigma_8$ & $S_8$ & $h$ \\
 \hline\hline
Uniform & $0.277\pm 0.029$ & $0.856\pm 0.100$ & $0.818\pm 0.070$ & $0.662\pm 0.023$ \\
HST & $0.318\pm 0.029$ & $0.756\pm 0.089$ & $0.775\pm 0.071$ & - \\
Planck & $0.285\pm 0.023$ & $0.833\pm 0.088$ & $0.809\pm 0.069$ & - \\
\hline
\end{tabular}
\caption{Mean and standard deviation of $\Omega_m$, $\sigma_8$, $S_8$ and $h$ from the joint analysis of the average sparsity and BAO data.}
\label{table_spars_bao}
\end{table}

\begin{table}[th]
\centering
Sparsity + Gas Mass Fraction \\
\begin{tabular}{ccccc}
\hline\hline
 $f_{\rm gas}$ and $h$ priors & $\Omega_m$ & $\sigma_8$ & $S_8$ & $h$\\
 \hline\hline
$Y_{b,500c}^{\rm The 300}$ + $K^{\rm CMB}_{500c}$ + Uniform & $0.13 \pm 0.01$ & $1.35\pm 0.18$ & $0.90\pm 0.09$ & $1.10\pm 0.05$\\
  $Y_{b,500c}^{\rm The 300}$ + $K^{\rm CLASH}_{500c}$ + Uniform & $0.19 \pm 0.03$ & $1.04\pm 0.14$ & $0.82\pm 0.08$ & $0.91\pm 0.06$\\
  $Y_{b,500c}^{\rm FABLE}$ + $K^{\rm CLASH}_{500c}$ + Uniform & $0.19 \pm 0.03$ & $1.06\pm 0.15$ & $0.83\pm 0.08$ & $0.92\pm 0.07$\\
 $Y_{b,500c}^{\rm The 300}$ + $K^{\rm CCCP}_{500c}$ + Uniform & $0.20 \pm 0.02$ & $1.03\pm 0.12$ & $0.83\pm 0.07$ & $0.89\pm 0.05$\\ 
 $Y_{b,500c}^{\rm The 300}$ $(K_{500c}=1)$ + Uniform & $0.25 \pm 0.03$ & $0.86\pm 0.10$ & $0.79\pm 0.07$ & $0.79\pm 0.05$\\
 \hline
 $Y_{b,500c}^{\rm The 300}$ + $K^{\rm CMB}_{500c}$ + HST & $0.16 \pm 0.01$ & $1.27\pm 0.04$ & $0.92\pm 0.03$ & - \\
  $Y_{b,500c}^{\rm The 300}$ + $K^{\rm CLASH}_{500c}$ + HST  & $0.26\pm 0.03$ & $0.85\pm 0.10$ & $0.79\pm 0.07$ & -\\
    $Y_{b,500c}^{\rm FABLE}$ + $K^{\rm CLASH}_{500c}$ + HST  & $0.26\pm 0.02$ & $0.87\pm 0.07$ & $0.81\pm 0.05$ & -\\
 $Y_{b,500c}^{\rm The 300}$ + $K^{\rm CCCP}_{500c}$ + HST  & $0.24\pm 0.02$ & $0.94\pm 0.11$ & $0.83\pm 0.08$ & - \\
 $Y_{b,500c}^{\rm The 300}$ $(K_{500c}=1)$ + HST  & $0.27\pm 0.03$ & $0.83\pm 0.09$ & $0.79\pm 0.07$ & -\\
 \hline
 $Y_{b,500c}^{\rm The 300}$ + $K^{\rm CMB}_{500c}$ + Planck  & $0.17 \pm 0.01$ & $1.25\pm 0.04$ & $0.94\pm 0.03$ & - \\
  $Y_{b,500c}^{\rm The 300}$ + $K^{\rm CLASH}_{500c}$ + Planck  & $0.30\pm 0.03$ & $0.78\pm 0.11$ & $0.78\pm 0.06$ & -\\
    $Y_{b,500c}^{\rm FABLE}$ + $K^{\rm CLASH}_{500c}$ + Planck  & $0.28\pm 0.01$ & $0.85\pm 0.06$ & $0.83\pm 0.01$ & -\\
 $Y_{b,500c}^{\rm The 300}$ + $K^{\rm CCCP}_{500c}$ + Planck & $0.23\pm 0.02$ & $0.99\pm 0.01$ & $0.87\pm 0.03$ & - \\
 $Y_{b,500c}^{\rm The 300}$ $(K_{500c}=1)$ + Planck & $0.31\pm 0.03$ & $0.74\pm 0.10$ & $0.75\pm 0.07$ & -\\
 \hline
\end{tabular}
\caption{Mean and standard deviation of $\Omega_m$, $\sigma_8$ and $S_8$ and $h$ from the joint analysis of the average sparsity and gas mass fraction data for different $Y_{b,500c}$, $K_{500,c}$ and $h$ priors. \label{tableparams}}
\end{table}

\begin{table}[th]
\centering
Sparsity + Gas Mass Fraction + BAO\\
\begin{tabular}{ccccc}
\hline\hline
 $h$-prior & $\Omega_m$ & $\sigma_8$ & $S_8$ & $h$ \\
 \hline\hline
Uniform & $0.316\pm 0.013$ & $0.757\pm 0.067$ & $0.776\pm 0.064$ & $0.696\pm 0.017$ \\
HST & $0.325\pm 0.012$ & $0.736\pm 0.066$ & $0.766\pm 0.065$ & - \\
Planck & $0.306\pm 0.010$ & $0.781\pm 0.061$ & $0.788\pm 0.059$ & - \\
\hline
\end{tabular}
\caption{Mean and standard deviation of $\Omega_m$, $\sigma_8$ and $S_8$ and $h$ from the joint analysis of the average sparsity, gas mass fraction and BAO data.}
\label{sparsfgasbao_tab}
\end{table}

%\section{Galaxy Cluster Sparsity and Gas Mass Fraction Data}\label{app5}

%\begin{deluxetable*}{ccccc}
%\tablecolumns{5}
%\tablewidth{0pt}
%\tabletypesize{12pt}
%\tablecaption{\label{tab:clusters}
%Galaxy Clusters from the {\it Selected} LC$^2$-{\rm single} and HSC-XXL catalogs}
%\tablehead{ Name & z & $M_{200c}$ & $M_{500c}$ & $s_{200,500}$ \\
%\colhead{} & \colhead{} & ($10^{14}\,M_{\odot}$) & ($10^{14}\,M_{\odot}$) & \colhead{} }
%\startdata
%0 & 0 & 0 & 0 & 0\\
%\enddata
%\tablecomments{xxx.}
%\tablenotetext{a}{XLSSC cluster identifier.}
%\end{deluxetable*}

\end{document}